\documentclass[a4paper,12pt]{article}

\usepackage[left=25mm, right=25mm, top=30mm, bottom=30mm]{geometry}
\usepackage[utf8]{inputenc}
\usepackage[english]{babel}
\usepackage{amsmath}
\usepackage{amsfonts}
\usepackage{mathrsfs}
\usepackage{dsfont}
\usepackage{graphicx}
\usepackage{cite}
\usepackage[colorlinks, allcolors=blue, linktocpage=true]{hyperref}

\renewcommand{\O}{\mathcal{O}}
\newcommand{\ket}[1]{\left| #1 \right\rangle}
\newcommand{\bra}[1]{\left\langle #1 \right|}
\newcommand{\SO}{\text{SO}}
\newcommand{\lambdabar}{\bar{\lambda}}
\DeclareMathOperator{\diag}{diag}
\renewcommand{\H}{\widehat{H}}
\newcommand{\D}{\widehat{D}}
\newcommand{\M}{\widehat{M}}
\newcommand{\K}{\widehat{K}}
\renewcommand{\k}{\widehat{k}}
\newcommand{\A}{\widehat{A}}
\newcommand{\Q}{\widehat{Q}}
\newcommand{\N}{\mathcal{N}}

\newcommand{\one}{{\, \rm I}}
\newcommand{\two}{{\, \rm II}}

\makeatletter
\newsavebox{\@brx}
\newcommand{\llangle}[1][]{\savebox{\@brx}{\(\m@th{#1\langle}\)}%
  \mathopen{\copy\@brx\kern-0.5\wd\@brx\usebox{\@brx}}}
\newcommand{\rrangle}[1][]{\savebox{\@brx}{\(\m@th{#1\rangle}\)}%
  \mathclose{\copy\@brx\kern-0.5\wd\@brx\usebox{\@brx}}}
\makeatother

\numberwithin{equation}{section}

\title{Spinors and conformal correlators}

\author{Marc Gillioz}

\date{\emph{SISSA, via Bonomea 265, 34136 Trieste, Italy}}

\begin{document} 
\maketitle

\begin{abstract}
In conformal field theory, momentum eigenstates can be parameterized by a pair of real spinors, in terms of which special conformal transformations take a simpler form. This well-known fact allows to express 2-point functions of primary operators in the helicity basis, exposing the consequences of unitarity. What is less known is that the same pair of spinors can be used, together with a pair of scalar quantities, to parameterize 3-point functions. We  develop this formalism in 3 dimensions and show that it provides a simple realization of the operator product expansion (OPE) for scalar primary operators acting on the vacuum.
\end{abstract}

\vfill

\tableofcontents

\vfill


\newpage
\section{Introduction}

Conformal field theory is remarkable in the sense that it admits a non-perturbative definition in terms of CFT data characterizing 2- and 3-point correlation functions only.
The problem of constructing all these 2- and 3-point functions in the position-space representation is in some sense completely solved: since conformal Ward identities are first-order differential equations in position space, it essentially reduces to the classification of all tensors transforming covariantly~\cite{Osborn:1993cr, Costa:2011mg, Costa:2014rya}.

A related problem is the construction of all 2- and 3-point correlation functions in the momentum-space representation. One could expect this problem to be solved as well, since the two representations are related by a Fourier transform. But this Fourier transform is very difficult to perform explicitly --- so difficult in fact that most results in momentum space do not rely on it in an essential way~\cite{Coriano:2013jba, Bzowski:2013sza}.

In this work we introduce a new approach for the construction of 2- and 3-point conformal correlation function in momentum space, making use of the spinor formalism. This method works specifically in Minkowski space-time and with Wightman functions, i.e.~with correlators that are not time-ordered.
Unlike other recent works that use similar methods~\cite{Caron-Huot:2021kjy, Jain:2021vrv}, we do not focus on (higher-spin) conserved currents but treat all conformal primary operators on the same footing.
To keep the argument simple, our results are limited to 3 space-time dimensions and, in the case of 3-point functions, to correlators involving two scalar operators. The generalization to more complicated correlators and to 4 space-time dimensions (or higher) will be presented elsewhere.
As a consequence, the correlators computed here are not new and can in fact be found in the literature~\cite{Gillioz:2018mto, Gillioz:2019lgs}.

Nevertheless, this work is more than an academic exercise: the use of spinors in Minkowski space-time allows to encode both the momentum and the polarization information of the operators~\cite{Mack:1969dg, Mack:1975je}, and therefore to test the Hilbert space structure of the theory, providing a simple understanding of unitarity bounds on the scaling dimensions of primary operators~\cite{Ferrara:1974pt, Mack:1975je, Minwalla:1997ka, Grinstein:2008qk}, and of reality conditions on OPE coefficients. Moreover, the method developed here will serve as a basis for future studies of generic conformal correlators involving primary operators carrying arbitrary spin representations.

Our results are presented as follows. Section~\ref{sec:2pt} describes the construction of 2-point functions. Generic CFT states carry massive momenta parameterized by a pair of real spinors.
We show that the Ward identity for special conformal transformations takes a simple form in the spinor formalism,
particularly when projected onto helicity eigenstates.
This is essentially a review of known facts, but it sets the stage for the computations of section~\ref{sec:3pt}. There, we show that the two independent momenta characterizing a 3-point function can be parameterized by the same two spinors, together with a pair of scalar quantities related to ratios of momenta squared.%
\footnote{This contrasts with standard on-shell methods for massive scattering amplitudes, in which \emph{every} massive momentum is parameterized by a pair of spinors~\cite{Arkani-Hamed:2017jhn}. In our case a \emph{single} pair of spinors is sufficient.}
Conformal Ward identities are derived and solved, yielding familiar results involving the Appell $F_4$ hypergeometric function, as well as a novel closed-form expression in the general case.
The results of sections~\ref{sec:2pt} and \ref{sec:3pt} are finally put together in section~\ref{sec:4pt}, leading to the formulation of an operator product expansion for scalar operators acting on the vacuum.
Some conclusions are drawn in section~\ref{sec:conclusions}.
Conventions for spinors in 3 space-time dimensions are given in appendix~\ref{sec:conventions}.


\section{2-point functions}
\label{sec:2pt}

We begin by considering states created by a single local primary operator acting on the vacuum, and work in the momentum-space representation such that
\begin{equation}
	P^\mu \O(p) \ket{0} = p^\mu \O(p) \ket{0},
	\label{eq:state}
\end{equation}
where $P^\mu$ is the generator of translations and $p^\mu$ its eigenvalue. 
$\ket{0}$ is the unique vacuum state of the theory, invariant under conformal transformations.
We shall also consider the inner product of such states, which by translation symmetry is proportional to a delta function,
\begin{equation}
	\bra{0} \O(p') \O(p) \ket{0}
	= (2\pi)^3 \delta^3(p' + p)
	\llangle \O(-p) \O(p) \rrangle.
	\label{eq:2pt}
\end{equation} 
The object inside double brackets is a function of a single 3-momentum $p^\mu$.

\subsection{Spinors and helicity eigenstates}

Standard assumptions of quantum field theory require states of the form of eq.~\eqref{eq:state} to only have support in the forward light cone.
In other words, the function in double bracket in eq.~\eqref{eq:2pt} vanishes unless
\begin{equation}
	p^2 = (p^0)^2 - \mathbf{p}^2 \geq 0,
	\qquad\qquad
	p^0 \geq 0.
	\label{eq:forwardlightcone}
\end{equation}
Any momentum of this sort can be parameterized by a pair of real spinors, or equivalently by a unique complex spinor $\lambda^a$ and its complex conjugate $\lambdabar^a$ through
\begin{equation}
	p^\mu = \sigma^\mu_{ab} \, \lambda^a \lambdabar^b.
	\label{eq:p}
\end{equation}
Note that this parameterization is redundant: it uses a complex spinor with 4 components (two real and two imaginary) for a real 3-vector.
This redundancy is encoded in the little group symmetry
\begin{equation}
	\lambda^a \to e^{i \theta} \lambda^a,
	\qquad\qquad
	\lambdabar^a \to e^{-i \theta} \lambdabar^a,
	\label{eq:littlegroup}
\end{equation}
under which $p$ transforms trivially. We say that $\lambda^a$ carries helicity $+\frac{1}{2}$ and $\lambdabar^a$ helicity $-\frac{1}{2}$.
In the rest frame in which $\mathbf{p} = 0$, this corresponds to spatial rotations. 

Spinors are particularly powerful when dealing with operators $\O$ that transform in non-trivial representations of the Lorentz group. In 3 space-time dimensions, all such operators are symmetric tensors in spinor indices. 
We shall denote with $2j$ the number of spinor indices, $j$ taking either integer or half-integer values. A complete and orthogonal basis of polarization tensors is given by
\begin{equation}
	\varepsilon^{a_1 \ldots a_{2j}}_m
	= \lambda^{a_1} \cdots \lambda^{a_{j+m}}
	\lambdabar^{a_{j+m+1}} \cdots \lambdabar^{a_{2j}},
	\label{eq:polarization}
\end{equation}
where $m$ takes values between $-j$ and $+j$ in integer steps.
There are $2j + 1$ such polarization tensors for each spin $j$, and all of them are eigenstates of helicity $m$, i.e.~they transform as $\varepsilon^{a_1 \ldots a_{2j}}_m \to e^{2i m \theta} \varepsilon^{a_1 \ldots a_{2j}}_m$ under the little group scaling \eqref{eq:littlegroup}.

To simplify the notation even further, we introduce an auxiliary spinor $\zeta^a$ and define the index-free polarization
\begin{equation}
	\varepsilon_{j,m}(\zeta)
	= (\zeta \cdot \lambda)^{j+m}
	(\zeta \cdot \lambdabar)^{j-m}.
	\label{eq:polarization:indexfree}
\end{equation}
The explicit form \eqref{eq:polarization} of the polarization tensor can be unequivocally recovered from this expression acting with spinor derivatives.%
\footnote{This contrasts with the use of polarization \emph{vectors} that require subtraction of trace terms~\cite{Costa:2011mg, Dobrev:1975ru}; there are no such trace terms in the spinor formalism.}
The same auxiliary spinor can be used to express any operator carrying spinor indices in the index-free notation
\begin{equation}
	\O_j(p, \zeta) \ket{0}
	\equiv \zeta^{a_1} \cdots \zeta^{a_{2j}}
	\O_{a_1 \ldots a_{2j}}(p) \ket{0}.
	\label{eq:state:spinning}
\end{equation}
The completeness of the basis of polarization tensors implies that we can further decompose this operator into helicity eigenstates as
\begin{equation}
	\O_j(p, \zeta) \ket{0}
	= \sum_{m = -j}^j
	\varepsilon_{j,m}(\zeta)
	\O_{j,m}(p) \ket{0}.
	\label{eq:helicityeigenstates}
\end{equation}
Note that these last two equations involve a primary operator $\O$ acting on the vacuum, as it is important to realize that they are only valid when $p$ takes values inside the forward light cone. The decomposition does not make sense at the level of operators.

\subsection{Conformal Ward identities}

The norm of these states is computed by the 2-point correlation function
\begin{equation}
	\bra{0} \O_j(p', \zeta') \O_j(p, \zeta) \ket{0}
	= (2\pi)^3 \delta^3(p' + p)
	\llangle \O_j(-p, \zeta') \O_j(p, \zeta) \rrangle,
	\label{eq:2pt:spinning}
\end{equation}
where we are now making use of two distinct auxiliary polarization spinors $\zeta$ and $\zeta'$ to encode the information about indices.
The object in double brackets is a correlation function in its own right: taking the Fourier transform of this equation with respect to $p'$, one gets 
\begin{equation}
	\llangle \O_j(-p, \zeta') \O_j(p, \zeta) \rrangle
	= e^{i p \cdot x}
	\bra{0} \widetilde{\O}_j(x, \zeta') \O_j(p, \zeta) \ket{0}.
	\label{eq:2pt:mixed}
\end{equation}
This mixed position/momentum representation is particularly well-suited to examine the constraints imposed by conformal symmetry: when evaluated at $x = 0$, the primary operator $\widetilde{\O}(0, \zeta')$ has trivial transformation properties under the conformal group. This implies that conformal Ward identities for this correlation function can be obtained in the form of differential equations with respect to the momentum $p^\mu$. For instance, for scale and Lorentz symmetry, we have
\begin{align}
	\left[ \D - 2\Delta + 3 \right]
	\llangle \O_j(-p, \zeta') \O_j(p, \zeta) \rrangle
	& = 0,
	\label{eq:WI:scale:2pt}
	\\[5mm]
	\left[ \M_{\mu\nu} - \widehat{\Sigma}_{\mu\nu}(\zeta')
	- \widehat{\Sigma}_{\mu\nu}(\zeta) \right]
	\llangle \O_j(-p, \zeta') \O_j(p, \zeta) \rrangle
	& = 0,
	\label{eq:WI:Lorentz:2pt:vector}
\end{align}
where
\begin{equation}
	\D = p^\mu \frac{\partial}{\partial p^\mu},
	\qquad\qquad
	\M_{\mu\nu} = p_\nu \frac{\partial}{\partial p^\mu}
	- p_\mu \frac{\partial}{\partial p^\nu}.
	\label{eq:D:M:p}
\end{equation}
$\Delta$ is the scaling dimension of $\O$, and $\widehat{\Sigma}_{\mu\nu}$ the spin operator acting on its Lorentz indices: in terms of the auxiliary spinor $\zeta$, it can be written as 
\begin{equation}
	\widehat{\Sigma}_{\mu\nu}(\zeta)
	= -\frac{1}{2} \varepsilon_{\mu\nu\rho} \sigma^\rho_{ab} \,
	\zeta^a \frac{\partial}{\partial \zeta_b}.
	\label{eq:spin}
\end{equation}
It is a simple exercise to verify using the relation \eqref{eq:p} and the chain rule for derivatives that the scale and Lorentz generators can be written in terms of spinors as
\begin{equation}
	\D = \frac{1}{2} \left( \lambda^a \frac{\partial}{\partial \lambda_a}
	+ \lambdabar^a \frac{\partial}{\partial \lambdabar_a} \right)
	\label{eq:D}
\end{equation}
and
\begin{equation}
	\M_{\mu\nu} = \frac{1}{2}
	\varepsilon_{\mu\nu\rho} \sigma^\rho_{ab} 
	\left( \lambda^a \frac{\partial}{\partial \lambda_b}
	+ \lambdabar^a \frac{\partial}{\partial \lambdabar_b} \right).
	\label{eq:M}
\end{equation}
These results can be obtained enumerating the handful of operators that are first-order in derivatives, invariant under the little group scaling~\eqref{eq:littlegroup}, and that transform respectively as a scalar and as a vector%
\footnote{In 3 dimensions $\M_{\mu\nu}$ is dual to a Lorentz vector, which is obvious in eq.~\eqref{eq:M}. We use the two-index antisymmetric form to adhere to the standard notation for conformal generators valid in arbitrary space-time dimension and to match the definition~\eqref{eq:D:M:p} in terms of $p^\mu$.}
under Lorentz symmetry.
Note that the polarization tensors \eqref{eq:polarization} are eigenvectors of the scale generator,
\begin{equation}
	\D \, \varepsilon_{j,m}(\zeta) = j \, \varepsilon_{j,m}(\zeta),
\end{equation}
and moreover that they transform covariantly under Lorentz transformation, since they satisfy
\begin{equation}
	\M_{\mu\nu} \varepsilon_{j,m}(\zeta)
	= \widehat{\Sigma}_{\mu\nu}(\zeta) \varepsilon_{j,m}(\zeta).
\end{equation}

The same strategy can be applied to determine the constraints imposed by special conformal symmetry, encoded in the Ward identity
\begin{equation}
	\K_\mu \llangle \O_j(-p, \zeta') \O_j(p, \zeta) \rrangle = 0,
	\label{eq:WI:special:2pt}
\end{equation}
where%
\footnote{The fact that eq.~\eqref{eq:K:2pt:p} does not depend on the auxiliary spinor $\zeta'$ and looks therefore asymmetric is due to the representation \eqref{eq:2pt:mixed} in which special conformal transformation act trivially on one of the two operators, but not on the other.
Alternatively, this Ward identity could be obtained commuting the generator of special conformal transformation with the delta function imposing momentum conservation~\cite{Maldacena:2011nz}.}
\begin{equation}
	\K_\mu = 
	\frac{1}{2} p_\mu \frac{\partial^2}{\partial p^\nu \partial p_\nu}
	- p^\nu \frac{\partial^2}{\partial p^\nu \partial p^\mu}
	+ (\Delta - 3) \frac{\partial}{\partial p^\mu}
	+ \widehat{\Sigma}_{\mu\nu}(\zeta) \frac{\partial}{\partial p_\nu}.
	\label{eq:K:2pt:p}
\end{equation}
In this case the expression for $\K_\mu$ in terms of spinors is more complicated because the Ward identity involves second-order derivatives in the momenta.
Nevertheless, it can be determined straightforwardly using
\begin{equation}
	\frac{\partial}{\partial p_\mu}
	= \frac{\sigma^\mu_{ab}}{4 \lambda \cdot \lambdabar}
	\left( \lambdabar^a \frac{\partial}{\partial \lambdabar_b}
	- \lambda^a \frac{\partial}{\partial \lambda_b} \right),
	\label{eq:dp}
\end{equation}
together with the identity
\begin{align}
	\frac{1}{2} p^\mu \frac{\partial^2}{\partial p^\nu \partial p_\nu}
	- p^\nu \frac{\partial^2}{\partial p^\nu \partial p_\mu}
	- \frac{\partial}{\partial p_\mu}
	= -\frac{1}{2} \sigma^\mu_{ab}
	\frac{\partial^2}{\partial \lambda_a \partial \lambdabar_b},
	\label{eq:ddp}
\end{align}
in which we note that the combination of second-order derivatives precisely matches that of eq.~\eqref{eq:K:2pt:p}.

This is sufficient to write down a differential equation for the correlation function in terms of spinors only.
But the effectiveness of the spinor formalism goes further, as it also implies that we can decompose this equation into 3 components of definite helicity, writing
\begin{equation}
	\K^\mu = \frac{\sigma^\mu_{ab}}{2 (\lambda \cdot \lambdabar)^2} 
	\left( \lambda^a \lambdabar^b \K_0
	+ \lambda^a \lambda^b \K_-
	+ \lambdabar^a \lambdabar^b \K_+ \right).
	\label{eq:K:projection}
\end{equation}
To give explicit expressions for the differential operators $\K_0$ and $\K_\pm$, it is convenient to construct all scalar, first-order derivative operators in spinors. There are four independent operators: one is the scale generator \eqref{eq:D}, the other three can be written
\begin{equation}
	\H_0 = \frac{1}{2} \left(
	\lambda^a \frac{\partial}{\partial \lambda^a}
	- \lambdabar^a \frac{\partial}{\partial \lambdabar^a} \right),
	\qquad\qquad
	\H_+ = \lambda^a \frac{\partial}{\partial \lambdabar^a},
	\qquad\qquad
	\H_- = \lambdabar^a \frac{\partial}{\partial \lambda^a}.
	\label{eq:H}
\end{equation}
$\H_0$ counts the helicity, while $\H_+$ and $\H_-$ are ladder operator that raise and lower the helicity by one unit. Together, they form an $\text{su}(2)$ algebra
\begin{equation}
	\left[ \H_0, \H_\pm \right] = \pm \H_\pm,
	\qquad\qquad
	\left[ \H_+, \H_- \right] = 2 \H_0.
	\label{eq:helicity:algebra}
\end{equation}
Note that $\H_+$ and $\H_-$ are complex-conjugate to each other, and also that all three commute with $\D$.
Any differential operator can then be expressed in terms of these: this is the case of the Lorentz generator
\begin{equation}
	\varepsilon^{\mu\nu\rho} \M_{\nu\rho}
	= \frac{\sigma^\mu_{ab}}{\lambda \cdot \lambdabar}
	\left( 2 \lambda^a \lambdabar^b \H_0
	+ \lambdabar^a \lambdabar^b \H_+
	- \lambda^a \lambda^b \H_- \right),
\end{equation}
and of the derivative with respect to $p$
\begin{equation}
	\frac{\partial}{\partial p_\mu}
	= \frac{\sigma^\mu_{ab}}{2 (\lambda \cdot \lambdabar)^2}
	\left( -2 \lambda^a \lambdabar^b \D
	+ \lambdabar^a \lambdabar^b \H_+
	+ \lambda^a \lambda^b \H_- \right).
	\label{eq:dp:projection}
\end{equation}
To express the components of $\K^\mu$ in this manner, one also needs to decompose the spin operator $\widehat{\Sigma}_{\mu\nu}$ into scalar components. This can be done defining
\begin{equation}
	\H^\zeta_+ =
	\frac{(\zeta \cdot \lambda) \lambda^a }
	{\lambda \cdot \lambdabar}
	\frac{\partial}{\partial \zeta^a},
	\qquad
	\H^\zeta_- = 
	- \frac{(\zeta \cdot \lambdabar) \lambdabar^a}
	{\lambda \cdot \lambdabar}
	\frac{\partial}{\partial \zeta^a},
	\qquad
	\H^\zeta_0 = - \frac{(\zeta \cdot \lambdabar) \lambda^a
	+ (\zeta \cdot \lambda) \lambdabar^a}{2 \, \lambda \cdot \lambdabar}
	\frac{\partial}{\partial \zeta^a}.
	\label{eq:H:zeta}
\end{equation}
These 3 operators obey the same $\text{su}(2)$ algebra as the $\H_i$, but act only on functions of the spinor $\zeta$. A set of differential operators acting on the auxiliary spinor $\zeta'$ can be defined similarly, denoted $\H^{\zeta'}_i$.
With these tools, the Ward identity \eqref{eq:WI:Lorentz:2pt:vector} for Lorentz transformations can be projected onto components of definite helicity, leading to the three simple equations
\begin{equation}
	\left[ \H_i - \H^\zeta_i - \H^{\zeta'}_i \right]
	\llangle \O_j(-p, \zeta') \O_j(p, \zeta) \rrangle
	= 0,
	\qquad\qquad
	i \in \{ 0, + , - \}.
	\label{eq:WI:Lorentz:2pt}
\end{equation}
The same can be done with the components of $\K_\mu$ defined in eq.~\eqref{eq:K:projection}. We find
\begin{align}
	\K_0 &= \H_+ \H_- - \H_+ \H^\zeta_- - \H_- \H^\zeta_+
	+ \D (\D - 2 \Delta + 3) - \H_0 (\H_0 + 1),
	\nonumber
	\\[5mm]
	\K_+ &= \H_+ ( -\D - \H_0 + \H_0^\zeta + \Delta - 2 )
	+ \H^\zeta_+ (\D + 1),
	\label{eq:K:2pt}
	\\[5mm]
	\K_- &= \H_- ( -\D + \H_0 - \H_0^\zeta + \Delta - 2 )
	+ \H^\zeta_- (\D + 1).
	\nonumber
\end{align}
This is a major improvement compared to the momentum-space Ward identity \eqref{eq:K:2pt:p}: on the one hand, the helicity projection disentangles the equation into three independent components; on the other hand, it is very simple to apply the elementary differential operators in eq.~\eqref{eq:K:2pt} to \emph{any} function of the spinors.
The scalar quantity $p^2 = -(\lambda \cdot \lambdabar)^2$ is for instance annihilated by all $\H_i$. Moreover, the polarization tensors \eqref{eq:polarization} have simple transformation rules, being eigenstates of $\H_0$,
\begin{equation}
	\H_0 \, \varepsilon_{j,m}(\zeta) = m \, \varepsilon_{j,m}(\zeta),
\end{equation}
and satisfying ladder identities
\begin{equation}
	\H_+ \, \varepsilon_{j,m}(\zeta)
	= (j - m) \, \varepsilon_{j,m+1}(\zeta),
	\qquad\qquad
	\H_- \, \varepsilon_{j,m}(\zeta)
	= (j + m) \, \varepsilon_{j,m-1}(\zeta).
	\label{eq:ladder}
\end{equation}
The numerical factors in these last two equations guarantee that $\H_\pm$ annihilate the highest-helicity polarizations $m = \pm j$ respectively.
The other differential operators $\H^\zeta_i$ and $\H^{\zeta'}_i$ act likewise, but only on polarization tensors defined in terms of the corresponding auxiliary spinor.


\subsection{Correlation function}

Now that the conformal Ward identities have been split into simple equations in terms of elementary ladder operators, it is straightforward to obtain the most general solution for the 2-point correlator \eqref{eq:2pt:spinning}.
Note that it is the overlap of two states that can each be decomposed into helicity eigenstates following eq.~\eqref{eq:helicityeigenstates}.
Moreover, since the correlation function~\eqref{eq:2pt:spinning} transform covariantly under the little group, its total helicity is zero. This implies that the decomposition into helicity eigenstates is diagonal, and that one can write
\begin{equation}
	\llangle \O_j(-p, \zeta') \O_j(p, \zeta) \rrangle
	= \sum_{m = -j}^j \varepsilon_{j, -m}(\zeta') \varepsilon_{j,m}(\zeta)
	\llangle \O_{j,-m}(-p) \O_{j,m}(p) \rrangle.
	\label{eq:2pt:helicity}
\end{equation}
$\llangle \O_{j,-m}(-p) \O_{j,m}(p) \rrangle$ is a scalar function of the spinors $\lambda^a$ and $\lambdabar^a$, which means that it must necessarily be a function of $p^2 = - (\lambda \cdot \lambdabar)^2$ only.
Its overall scaling dimension is determined by the Ward identity \eqref{eq:WI:scale:2pt} to be $2 \Delta - 2j - 3$.
We obtain immediately that 
\begin{equation}
	\llangle \O_{j,-m}(-p) \O_{j,m}(p) \rrangle
	= (p^2)^{\Delta - j - 3/2} C_{\Delta, j, m},
	\label{eq:2pt:ansatz}
\end{equation}
where $C_{\Delta, j, m}$ is a numerical coefficient that remains to be determined. Note that unitarity requires this coefficient to be non-negative.
Together, eqs.~\eqref{eq:2pt:helicity} and \eqref{eq:2pt:ansatz} form the most general ansatz for the 2-point function that is consistent with Poincar\'e and scale symmetry, i.e.~that satisfies the Ward identities \eqref{eq:WI:scale:2pt} and \eqref{eq:WI:Lorentz:2pt}.

In fact, the zero-helicity component $\K_0$ of the special conformal Ward identity~\eqref{eq:WI:special:2pt} is also readily satisfied by this ansatz.
This is not the case of the components $\K_\pm$, however:
using
\begin{align}
	\K_+ & \left[ (p^2)^{\Delta - j - 3/2}
	\varepsilon_{j, -m}(\zeta') \varepsilon_{j,m}(\zeta) \right]
	= (p^2)^{\Delta - j - 3/2} 
	\nonumber \\
	&\times \left[ (j + m) (\Delta - m - 1)
	\varepsilon_{j, -m}(\zeta') \varepsilon_{j,m-1}(\zeta)
	- (j - m) (\Delta + m - 1)
	\varepsilon_{j, -m-1}(\zeta') \varepsilon_{j,m}(\zeta) \right],
\end{align}
or a similar identity for $\K_-$ related to this one by complex conjugation,
one obtains the condition
\begin{align}
	\sum_{m = -j}^{j-1} &
	\varepsilon_{j, -m-1}(\zeta') \varepsilon_{j,m}(\zeta)
	\nonumber \\
	& \times
	\left[ (j + m + 1) (\Delta - m - 2) C_{\Delta, j, m+1}
	- (j - m) (\Delta + m -1) C_{\Delta,j,m} \right] = 0.
\end{align}
For the Ward identity to be satisfied, all individual terms in the sum must vanish separately. This gives a recursion relation for the coefficient $C_{\Delta, j, m}$, solved by
\begin{equation}
	C_{\Delta,j,m} = \frac{(2j)!}{(j-m)! (j+m)!}
	\frac{\Gamma(\Delta + m - 1) \Gamma(\Delta - m - 1)}
	{\Gamma(\Delta + j - 1) \Gamma(\Delta - j - 1)} \, C_{\Delta, j}.
	\label{eq:C}
\end{equation}
$C_{\Delta, j}$ is a normalization constant that is arbitrary up to its sign:
since $C_{\Delta, j, \pm j} = C_{\Delta, j}$ corresponds to the norm of the eigenstate of helicity $\pm j$, it must therefore be positive.
Despite the appearance of gamma functions in eq.~\eqref{eq:C}, the $C_{\Delta, j, m}$ are in fact rational functions of $\Delta$ multiplying this normalization constant: they can alternatively be written using the Pochhammer symbol as
\begin{equation}
	C_{\Delta,j,m} = \frac{(2j)!}{(j-m)! (j+m)!}
	\frac{(\Delta - j - 1)_{j-m}}
	{(\Delta + m - 1)_{j-m}} \, C_{\Delta, j}.
\end{equation}
Requiring that all helicity eigenstates have positive norm,
one obtains the unitarity bound
\begin{equation}
	\Delta \geq j + 1
	\qquad\qquad\text{for all} \quad
	j > \tfrac{1}{2}.
\end{equation}
This is a well-known result, particularly transparent in this formalism.
When the bound is saturated, i.e.~when $\Delta = j + 1$, the operator only admits transverse polarizations with helicity $m = \pm j$: it is then a conserved current transforming in a short representation of the conformal group.%
\footnote{The conservation condition follows simply from the identity
$p_{a_1a_2} \varepsilon^{a_1 \ldots a_{2j}}_{\pm j} = 0$.}
In all other cases the operator admits $2j + 1$ polarizations, both transverse and longitudinal, and there is no shortening condition.

There are also well-known unitarity bounds for scalar ($j = 0$) and spinor ($j = \frac{1}{2}$) operators, but their origin is somewhat different. One way to understand them is to normalize the operator such that the position-space correlation function obeys, for space-like separated points,
\begin{equation}
	\bra{0} \widetilde{\O}_j(x', \zeta')
	\widetilde{\O}_j(x, \zeta) \ket{0}
	= \frac{\left[ i \sigma^\mu_{ab} \zeta'^a \zeta^b (x'_\mu - x_\mu) \right]^{2j}}
	{\left[ -(x' - x)^2 \right]^{\Delta + j}},
	\qquad\quad
	(x' - x)^2 < 0.
\end{equation}
Performing the Fourier transform of this expression, one recovers the 2-point correlation function \eqref{eq:2pt:spinning} with normalization%
\footnote{The case of integer spin is treated in ref.~\cite{Gillioz:2018mto}. For half-integer spins, we refer the reader to the excellent discussion of ref.~\cite{Erramilli:2019njx}.}
\begin{equation}
	C_{\Delta, j} = \frac{2 \pi^2}
	{(\Delta + j - 1) \Gamma(2\Delta - 2)}.
\end{equation}
This is positive for all $\Delta \geq \frac{1}{2}$ for a scalar, and for all $\Delta \geq 1$ for a spin-$\frac{1}{2}$ operator.

\paragraph{Distinct operators.}

We focused so far on 2-point functions of identical operators, but the methods presented in this section can also be used to examine the correlation function of two distinct operators. Based on Poincar\'e and scale symmetry, we can make the following ansatz:
\begin{equation}
	\llangle \O'_{j'}(-p, \zeta') \O_j(p, \zeta) \rrangle
	\stackrel{?}{=}
	\sum_m
	C_{\Delta, j', j, m}
	\varepsilon_{j', -m}(\zeta') \varepsilon_{j,m}(\zeta)
	(p^2)^{(\Delta' + \Delta - j' - j - 3)/2}.
	\label{eq:2pt:distinct}
\end{equation}
$\Delta'$ and $\Delta$ are the scaling dimensions of $\O'$ and $\O$ respectively. The sum over $m$ runs over $\left| m \right| \leq \min(j', j)$, and in this way invariance under the little group is also satisfied.
When it comes to special conformal transformation, however, this ansatz is constrained further: the zero-helicity component $\K_0$ of the Ward identity \eqref{eq:WI:special:2pt} now gives
\begin{equation}
\begin{aligned}
	\K_0 & \llangle \O'_{j'}(-p, \zeta') \O_j(p, \zeta) \rrangle
	\\
	& = \big[ (\Delta' - \Delta) (\Delta' + \Delta - 3)
	+ (j' - j) (j' + j + 1) \big]
	\llangle \O'_{j'}(-p, \zeta') \O_j(p, \zeta) \rrangle.
\end{aligned}
\end{equation}
For identical spins $j' = j$, special conformal symmetry requires that the two operators have either identical scaling dimension $\Delta' = \Delta$, or that their scaling dimensions add up as $\Delta' + \Delta = 3$. This latter condition is only compatible with the unitarity bounds for scalar operators.
When the spins are not equal, $j' \neq j$, the other Ward identities generated by $\K_\pm$ require the highest-helicity coefficient with $m = \pm \min(j', j)$ to vanish, and all other coefficients must therefore vanish as well (again assuming that the unitarity bounds are enforced).
This means that 2-point functions of distinct operators generically vanish in conformal field theory.%
\footnote{A possible exception concerns scalar operators with scaling dimensions adding up to 3, which is usually not realized in physical theories. The other exception concerns theories that have multiple operators in the same conformal representation: it is usually possible to diagonalize the basis of such operators, but not always~\cite{Hogervorst:2016itc}.
We shall not consider this possibility further.}

This concludes the discussion on 2-point correlations functions. The results obtained here are not new: they have recently appeared in the bootstrap literature as an efficient tool for the construction of conformal blocks in position space~\cite{Erramilli:2019njx}, and a similar construction in 4 space-time dimensions dates back to early days of conformal field theory half a century ago~\cite{Mack:1969dg, Mack:1975je}.%
\footnote{See also refs.~\cite{Henning:2019mcv, Henning:2019enq} for a recent use of the spinor-helicity formalism combined with conformal symmetry in the context of free theories.}
The probable reason why this approach is absent from modern reviews on the topic is that it is too remote from the position-space representation in which the OPE is usually formulated, and in terms of which conformal blocks can be simply expressed as functions of invariant cross-ratios.
The goal of this paper is precisely to show that the spinor formalism can be carried through the whole construction of conformal blocks, and that it is a valuable tool in doing so.


\section{3-point functions}
\label{sec:3pt}

Let us consider next 3-point correlation functions, specifically those involving two scalar operators, which we will denote by $\phi_1$ and $\phi_2$, and one arbitrary tensor operator $\O$ (only integer-spin operators will in fact appear, as explained below). As before, translation symmetry implies that we can write
\begin{equation}
	\bra{0} \O_{a_1 \ldots a_{2j}}(-p)
	\phi_2(p_2) \phi_1(p_1) \ket{0}
	= (2\pi)^3 \delta^3(p - p_2 - p_1)
	\llangle \O_{a_1 \ldots a_{2j}}(-p)
	\phi_2(p_2) \phi_1(p_1) \rrangle.
	\label{eq:3pt}
\end{equation}
One way to understand this Wightman function is as the overlap of two states, one obtained as the product of two operators acting successively on the vacuum,
\begin{equation}
	\phi_2(p_2) \phi_1(p_1) \ket{0},
	\label{eq:state:in}
\end{equation}
and the other of the form discussed in the previous section,
\begin{equation}
	\O_{a_1 \ldots a_{2j}}(p_1 + p_2) \ket{0}.
	\label{eq:state:out}
\end{equation}
Both are momentum-eigenstates, with eigenvalue $p = p_1 + p_2$ that we can assume to be in the forward light cone. We will also see that both can be expanded into helicity eigenstates: for the state \eqref{eq:state:out} the helicity labels the spin of the operator, carried by its spinor indices, while for the state \eqref{eq:state:in} it corresponds to angular momentum. 
Before discussing the constraints of conformal symmetry that will determine this 3-point function up to a single OPE coefficient, we need to introduce a unified language that can account for these two distinct manifestations of helicity.

\subsection{From momenta to spinors}

Once the delta function imposing momentum conservation has been taken out of its definition, the correlator in eq.~\eqref{eq:3pt} is a function of two 3-momenta $p_1$ and $p_2$ only.
The spectral condition on their sum is such that one can always parameterize it with a complex spinor as
\begin{equation}
	p^\mu = p_1^\mu + p_2^\mu 
	= \sigma^\mu_{ab} \, \lambda^a \lambdabar^b,
	\label{eq:p:sum}
\end{equation}
in complete analogy with eq.~\eqref{eq:p}. However, the 3-point function does not only depend on the sum $p_1 + p_2$, but also on the difference $p_1 - p_2$. Depending on the kinematics, this difference can either be space-like, time-like, or even light-like. It is therefore not immediately well-suited to be expressed in terms of spinors.
Instead, let us consider the vector%
\footnote{In the particle-physics literature, and particularly in the context of bound states of 2-particles, $q$ is sometimes called the \emph{relative momentum in the center-of-mass frame}. It is purely spatial ($q^0 = 0$) in a Lorentz frame in which $\mathbf{p}_1 + \mathbf{p}_2 = 0$. Moreover, if the ``constituent masses'' $p_1^2$ and $p_2^2$ are equal, then $\mathbf{q} \propto \mathbf{p}_1 - \mathbf{p}_2$. Eq.~\eqref{eq:q:def} is simply the Lorentz-covariant definition of this quantity.}
\begin{equation}
	q^\mu = \frac{\big[ (p_1 + p_2) \cdot p_2 \big] p_1^\mu 
	- \big[ (p_1 + p_2) \cdot p_1 \big] p_2^\mu}
	{\sqrt{(p_1 \cdot p_2)^2 - p_1^2 p_2^2}}.
	\label{eq:q:def}
\end{equation}
This is the unique combination of $p_1$ and $p_2$ (up to a sign) that is orthogonal to $p$, i.e.~$p \cdot q = 0$, and that has the same norm in absolute value, $q^2 = -p^2$ ($q$ is always space-like).
These two conditions are sufficient to give it a parameterization in terms of the same complex spinor:
\begin{equation}
	q^\mu = \tfrac{1}{2} \sigma^\mu_{ab}
	\left( \lambda^a \lambda^b + \lambdabar^a \lambdabar^b \right).
	\label{eq:q:spinors}
\end{equation}
$q$ is a real vector, but unlike $p$, it is not invariant under the little group \eqref{eq:littlegroup}. 

\begin{figure}
	\centering
	\includegraphics[width=10cm]{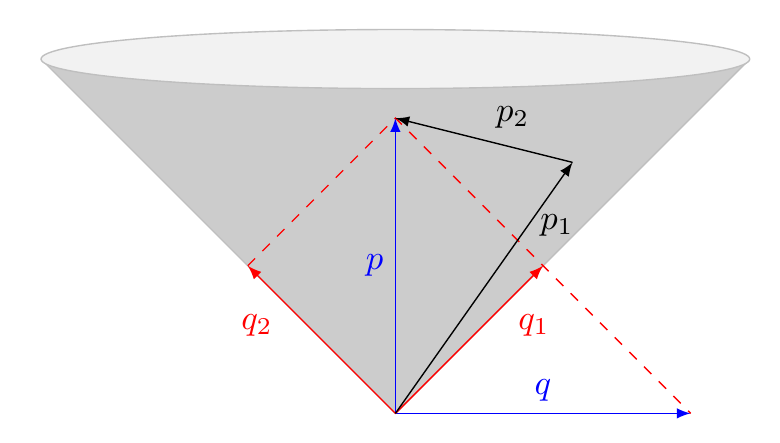}
	\caption{Given a pair of momenta $p_1$ and $p_2$ that add up to $p$
	inside the forward light cone, shown here in the center-of-mass
	frame ($\mathbf{p} = 0$), one can always define two null momenta
	$q_1$ and $q_2$ that are the projections of $p$ onto the light cone
	in the plane spanned by $p_1$ and $p_2$.
	The space-like vector $q = q_1 - q_2$
	satisfies $p \cdot q = 0$ and $q^2 = -p^2$.}
	\label{fig:momenta}
\end{figure}
A different perspective of the same parameterization goes as follows: there is a unique pair of null vectors $q_1$ and $q_2$ pointing in the forward direction that is formed of linear combinations of $p_1$ and $p_2$ and adds up to $p$, namely (up to the exchange of the two)
\begin{equation}
\begin{aligned}
	q_1^\mu &=
	\frac{1}{2} \left( 1 + \frac{p_2 \cdot (p_1 + p_2)}
	{\sqrt{(p_1 \cdot p_2)^2 - p_1^2 p_2^2}} \right) p_1^\mu 
	+ \frac{1}{2} \left( 1 - \frac{p_1 \cdot (p_1 + p_2)}
	{\sqrt{(p_1 \cdot p_2)^2 - p_1^2 p_2^2}} \right) p_2^\mu,
	\\
	q_2^\mu &=
	\frac{1}{2} \left( 1 - \frac{p_2 \cdot (p_1 + p_2)}
	{\sqrt{(p_1 \cdot p_2)^2 - p_1^2 p_2^2}} \right) p_1^\mu 
	+ \frac{1}{2} \left( 1 + \frac{p_1 \cdot (p_1 + p_2)}
	{\sqrt{(p_1 \cdot p_2)^2 - p_1^2 p_2^2}} \right) p_2^\mu.
\end{aligned}
\end{equation}
This is illustrated in figure~\ref{fig:momenta}.
Each of these two null vectors can then be written using a real spinor,
\begin{equation}
	q_1^\mu = \tfrac{1}{2} \sigma^\mu_{ab} \lambda_1^a \lambda_1^b,
	\qquad\qquad
	q_2^\mu = \tfrac{1}{2} \sigma^\mu_{ab} \lambda_2^a \lambda_2^b.
\end{equation}
Both the parameterization \eqref{eq:p:sum} for $p = q_1 + q_2$ and \eqref{eq:q:spinors} for $q = q_1 - q_2$ are recovered upon identifying $\lambda^a = \frac{1}{\sqrt{2}} \left( \lambda_1^a + i \lambda_2^a \right)$. Note that this procedure does not depend whether $p_1$ and $p_2$ are space-like or time-like
(even though the spectral condition also requires $p_1$ to be in the forward light cone).
The only requirement is that the two vectors are not colinear, so that 
$(p_1 \cdot p_2)^2 - p_1^2 p_2^2 > 0$.

It is important to realize that $q_1$ and $q_2$ are \emph{projections} of $p_1$ and $p_2$ onto the light cone, and as such the map is not invertible: to reconstruct $p_1$ and $p_2$ from of the massless vectors $q_1$ and $q_2$, we also need information about their length. 
To keep track of these, we introduce the dimensionless scalar quantities
$x$ and $y$, defined by the equations
\begin{equation}
	x (1 - y) = \frac{p_1^2}{(p_1 + p_2)^2},
	\qquad\qquad
	(1 - x) y = \frac{p_2^2}{(p_1 + p_2)^2},
	\label{eq:xy}
\end{equation}
in terms of which we can write
\begin{equation}
\begin{aligned}
	p_1^\mu &= \tfrac{1}{2} \sigma^\mu_{ab}
	\left[ (1 + x - y) \lambda^a \lambdabar^b
	+ \tfrac{1}{2} \, (1 - x - y)
	\left( \lambda^a \lambda^b + \lambdabar^a \lambdabar^b \right)
	\right],
	\\
	p_2^\mu &= \tfrac{1}{2}  \sigma^\mu_{ab}
	\left[ (1 + x - y) \lambda^a \lambdabar^b 
	- \tfrac{1}{2} \, (1 - x - y)
	\left( \lambda^a \lambda^b + \lambdabar^a \lambdabar^b \right) 
	\right].
\end{aligned}
\label{eq:p1:p2:lambda}
\end{equation}
This is the central observation of this work: any 3-point function can be parameterized with a \emph{unique} complex spinor, like the 2-point function, together with a pair of scalar quantities. The map
\begin{equation}
	\left\{ \lambda^a, \lambdabar^a, x, y \right\}
	\qquad\to\qquad
	\left\{ p_1^\mu, p_2^\mu \right\}
	\label{eq:map}
\end{equation}
is surjective, with both sides described by six real parameters. It is also regular as long as $p_1$ and $p_2$ are not colinear, i.e.~for $(p_1 \cdot p_2)^2 - p_1^2 p_2^2 = \frac{1}{4} (\lambda \cdot \lambdabar)^2 (1 - x - y)^2 \neq 0$.
If we restrict the range of $x$ and $y$ to the half-plane $x + y < 1$, then the map \eqref{eq:map} is in fact one-to-one.%
\footnote{The parameterization in terms of $x$ and $y$ provides in fact a double cover of the real kinematic domain in terms of $p_1^2/(p_1 + p_2)^2$, $p_2^2/(p_1 + p_2)^2$, but it has the advantage of avoiding the square root that is otherwise present in the definition \eqref{eq:q:def} of $q$.}
The problem of expressing the conformal Ward identities in this new set of parameters can therefore essentially be reduced to the computation of a Jacobian.

Before moving forward, let us emphasize once more that this construction is very specific to Minkowski space-time (spinors parameterize null vectors), and also to Wightman functions (the spectral condition is needed to guarantee that at least one momentum sits inside the forward light cone). It is not at all obvious how this could be extended to Euclidean space, or even to time-ordered correlation functions in Minkowski space-time.

\subsection{Conformal symmetry}

One of the great features of the map \eqref{eq:map} is that the two quantities $x$ and $y$ are \emph{scalar} and \emph{dimensionless}.
Only the spinor $\lambda^a$ and its complex conjugate are dimensionful and transform under Lorentz symmetry. This implies that it is straightforward to construct objects that transforms covariantly under these symmetries, up to arbitrary functions of $x$ and $y$.

In particular, all dependence on spinor indices can be expressed in terms of the polarization tensors~\eqref{eq:polarization}, or their index-free form \eqref{eq:polarization:indexfree}. This also implies that the decomposition \eqref{eq:helicityeigenstates} into helicity eigenstates is still valid, and that one can write
\begin{equation}
	\llangle \O_j(-p, \zeta) \phi_2(p_2) \phi_1(p_1) \rrangle
	= \sum_{m = -j}^j \varepsilon_{j,-m}(\zeta)
	\llangle \O_{j,m}(-p) \phi_2(p_2) \phi_1(p_1) \rrangle,
	\label{eq:3pt:helicity}
\end{equation}
where the correlator on the right-hand side is now a function of scalar quantities only, namely $x$, $y$ and $\lambda \cdot \lambdabar$, equivalent to the momenta-squared $p_1^2$, $p_2^2$ and $p^2$.
Moreover, since the overall scaling dimension of the correlator is known, we can further deduce that 
\begin{equation}
	\llangle \O_{j,m}(-p) \phi_2(p_2) \phi_1(p_1) \rrangle
	= (p^2)^{(\Delta_1 + \Delta_2 + \Delta - j - 6)/2}
	F_{j,m}(x, y),
	\label{eq:3pt:ansatz}
\end{equation}
where $\Delta_1$, $\Delta_2$ and $\Delta$ are respectively the scaling dimensions of $\phi_1$, $\phi_2$ and $\O$, and $F_{j,m}(x, y)$ is an function that is not constrained by Poincar\'e or scale symmetry.
This ansatz satisfies the Ward identities%
\footnote{Note that these identities involve the same generators \eqref{eq:D} and \eqref{eq:M} as in section~\ref{sec:2pt}.
They can be expressed in terms of the momenta $p_1$ and $p_2$ as shown in eq.~\eqref{eq:D:M:p1p2}.}
\begin{align}
	\left[ \M_{\mu\nu} - \widehat{\Sigma}_{\mu\nu}(\zeta) \right]
	\llangle \O_j(-p_1-p_2, \zeta) \phi_2(p_2) \phi_1(p_1) \rrangle
	& = 0,
	\label{eq:WI:Lorentz:3pt}
	\\[3mm]
	\left[ \D - \Delta - \Delta_1 - \Delta_2 + 6 \right]
	\llangle \O_j(-p_1-p_2, \zeta) \phi_2(p_2) \phi_1(p_1) \rrangle
	& = 0.
	\label{eq:WI:scale:3pt}
\end{align} 
In fact, each term in the sum \eqref{eq:3pt:helicity} individually satisfies these identities.

On the contrary, the Ward identity for special conformal transformations relates different helicity eigenstates, as we saw in section~\ref{sec:2pt},
and it also constrains the functions of $x$ and $y$ entering the ansatz~\eqref{eq:3pt:ansatz}.
In terms of the momenta $p_1$ and $p_2$, the correlation functions must satisfy the second-order differential equations
\begin{equation}
	\K^\mu \llangle \O_j(-p_1 - p_2, \zeta) \phi_2(p_2) \phi_1(p_1) \rrangle
	= 0,
	\label{eq:WI:special:3pt}
\end{equation}
where
\begin{equation}
	\K^\mu = 
	\sum_{i = 1}^2 \left[
	\frac{1}{2} p_i^\mu 
	\frac{\partial^2}{\partial p_i^\nu \partial p_{i\nu}}
	- p_i^\nu \frac{\partial^2}{\partial p_i^\nu \partial p_{i\mu}} 
	+ (\Delta_i - 3) \frac{\partial}{\partial p_{i\mu}} \right].
	\label{eq:K:3pt:p}
\end{equation}
Note that this equation does not depend on the scaling dimension $\Delta$ or spin $j$ of the operator $\O$: as with the 2-point function~\eqref{eq:2pt:mixed}, translation symmetry can be used to place $\O$ at the origin of the space-time coordinate system where special conformal transformations act trivially.
Nevertheless, solving this Ward identity is not an easy task.
Working with spinors once again proves useful: the 3 components of the vector operator $\K^\mu$ can be equivalently encoded into 3 scalar operators of helicity $-1$, $0$ and $+1$, as in eq.~\eqref{eq:K:projection}.
After computing the Jacobian of the map \eqref{eq:map} (details are given in appendix~\ref{sec:conventions}), one finds
\begin{equation}
\begin{aligned}
	\K_0 &= 2 \A^\one_{\Delta_1, \Delta_2; 6- \Delta_1 - \Delta_2, 0}
	- \frac{(1 - x + y) (1 + x - y)}{(1 - x - y)^2} \, \H_0^2
	+ \frac{1}{2} ( \H_+ \H_- + \H_- \H_+ )
	\\
	& \quad
	+ \D^2 + (9 - 2 \Delta_1 - 2 \Delta_2) \D
	\\
	\K_+ &= \A^\two_{\Delta_1, \Delta_2}
	- \H_+ ( \D + \H_0 + 5 - \Delta_1 - \Delta_2) 
	\\
	& \quad
	- \frac{1}{1 - x - y} \bigg[
	2 x (1 - x) \frac{\partial}{\partial x}
	- \left( \Delta_1 - 2 - \tfrac{1}{2} \H_0 \right) (1 - x + y)
	\\
	& \hspace{30mm}
	- 2 y (1 - y) \frac{\partial}{\partial y}
	+ \left( \Delta_2 - 2 - \tfrac{1}{2} \H_0 \right) (1 + x - y)
	\bigg] \H_0
	\\
	\K_- &= \A^\two_{\Delta_1, \Delta_2}
	- \H_- ( \D - \H_0 + 5 - \Delta_1 - \Delta_2) 
	\\
	& \quad
	+ \frac{1}{1 - x - y} \bigg[
	2 x (1 - x) \frac{\partial}{\partial x}
	- \left( \Delta_1 - 2 + \tfrac{1}{2} \H_0 \right) (1 - x + y)
	\\
	& \hspace{30mm}
	- 2 y (1 - y) \frac{\partial}{\partial y}
	+ \left( \Delta_2 - 2 + \tfrac{1}{2} \H_0 \right) (1 + x - y)
	\bigg] \H_0
\end{aligned}
	\label{eq:K:3pt}
\end{equation}
where we have defined a pair of second-order differential operators in $x$ and $y$
\begin{align}
	\A^\one_{\Delta_1, \Delta_2; \Delta, m} &= 
	x (1 - x) \frac{\partial^2}{\partial x^2}
	+ y (1 - y) \frac{\partial^2}{\partial y^2}
	\nonumber \\
	& \quad
	+ \left[ \left( \tfrac{5}{2} - \Delta_1 \right) (1 - x)
	- \left( \tfrac{5}{2} - \Delta_2 \right) x
	- (2m + 1) \frac{x (1-x)}{1 - x - y} \right]
	\frac{\partial}{\partial x}
	\nonumber \\
	& \quad
	+ \left[ \left( \tfrac{5}{2} - \Delta_2 \right) (1 - y)
	- \left( \tfrac{5}{2} - \Delta_1 \right) y
	- (2m + 1) \frac{y (1-y)}{1 - x - y} \right]
	\frac{\partial}{\partial y}
	\nonumber \\
	& \quad
	- \frac{1}{2} (6 - \Delta_1 - \Delta_2 - \Delta + m)
	(3 - \Delta_1 - \Delta_2 + \Delta + m)
	\label{eq:A:1}
\end{align}
and
\begin{equation}
\begin{aligned}
	\A^\two_{\Delta_1, \Delta_2}
	&= x (1 - x) \frac{\partial^2}{\partial x^2}
	+ \left[ \left( \tfrac{5}{2} - \Delta_1 \right) (1 - x)
	- \left( \tfrac{5}{2} - \Delta_2 \right) x \right]
	\frac{\partial}{\partial x}
	\\
	& \quad
	- y (1 - y) \frac{\partial^2}{\partial y^2}
	- \left[ \left( \tfrac{5}{2} - \Delta_2 \right) (1 - y)
	- \left( \tfrac{5}{2} - \Delta_1 \right) y \right]
	\frac{\partial}{\partial y}.
\end{aligned}
\label{eq:A:2}
\end{equation}
These are differential operators of the Appell $F_4$ type (see below).
A remarkable property of the $\K_i$ is that they take a compact form when applied to eigenfunctions of the scale and helicity operators $\D$ and $\H_0$: using the fact that $\D = \Delta_1 + \Delta_2 + \Delta - 6$ and $\H_0 = m$ for each term in the ansatz \eqref{eq:3pt:helicity}, one can write 
\begin{align}
	\K_0 &= 2 (1 - x - y)^{\pm m}
	\A^\one_{\Delta_1, \Delta_2; \Delta, \pm m} (1 - x - y)^{\mp m} 
	+ \frac{1}{2} ( \H_+ \H_- + \H_- \H_+ )
	\mp m ,
	\nonumber \\
	\K_+ &= (1 - x - y)^{-m} \A^\two_{\Delta_1, \Delta_2} (1 - x - y)^m
	- (\Delta + m - 1) \H_+ ,
	\label{eq:K:3pt:eigenvalues}
	\\
	\K_- &= (1 - x - y)^m \A^\two_{\Delta_1, \Delta_2} (1 - x - y)^{-m}
	- (\Delta - m - 1) \H_- ,
	\nonumber 
\end{align}
where for $\K_0$ either sign is possible.
It is important to note that the differential operators $\A^{\one,\two}$ act to the right, meaning that they do not commute with the power of $(1-x-y)$.
As a reminder, $\H_\pm$ raise or lower the helicity of the polarization tensors $\varepsilon_{j,m}(\zeta)$, as in eq.~\eqref{eq:ladder}, and it acts trivially on $p^2 = - (\lambda \cdot \lambdabar)^2$. 
This implies that the Ward identity~\eqref{eq:WI:special:3pt} yields the following conditions on the ansatz \eqref{eq:3pt:ansatz}:
one set of equations involving the function $F_{j, m}(x,y)$ only,
\begin{equation}
	\A^\one_{\Delta_1, \Delta_2; \Delta, \pm m}
	\left[ (1 - x - y)^{\mp m} F_{j,m}(x, y) \right]
	= \frac{(j \mp m) (j \pm m + 1)}{2} \,  
	(1 - x - y)^{\mp m} F_{j,m}(x, y)
	\label{eq:Fjm:equation}
\end{equation}
and two recursion relations involving functions with consecutive values of $m$,
\begin{equation}
	\A^\two_{\Delta_1, \Delta_2}
	\left[ (1 - x - y)^{\mp m} F_{j,m}(x, y) \right]
	= (j \pm m + 1) (\Delta \mp m - 2)
	(1 - x - y)^{\mp m}  F_{j,m \pm 1}(x, y).
	\label{eq:Fjm:ladder}
\end{equation}
These equations apply to all functions $F_{j,m}$ with $m$ between $-j$ and $+j$. There are no functions with $|m| = j + 1$ in the sum \eqref{eq:3pt:helicity}: when $m = + j$ or $m = -j$, instead of eq.~\eqref{eq:Fjm:ladder}, one gets
\begin{equation}
	\A^\two_{\Delta_1, \Delta_2}
	\left[ (1-x-y)^{\mp j} F_{j,\pm j}(x, y) \right] = 0.
	\label{eq:Fjj:two}
\end{equation}
In the same cases, the right-hand side of eq.~\eqref{eq:Fjm:equation} vanishes with the appropriate choice of sign, and thus
\begin{equation}
	\A^\one_{\Delta_1, \Delta_2; \Delta, \pm j}
	\left[ (1-x-y)^{\mp j} F_{j,\pm j}(x, y) \right] = 0.
	\label{eq:Fjj:one}
\end{equation}
Eqs.~\eqref{eq:Fjm:equation} to \eqref{eq:Fjj:one} incorporate all the constraints that conformal symmetry imposes on the 3-point correlator. Eq.~\eqref{eq:Fjm:ladder} is particularly important as it relates all functions $F_{j,m}$ with distinct $m$ in an unequivocal way: if any of these functions is known, then all the others can be obtained recursively applying the simple differential operator \eqref{eq:A:2}.

\subsection{Solutions}

In addition to this recursion relation, the two equations \eqref{eq:Fjj:two} and \eqref{eq:Fjj:one} taken together are \emph{nearly} sufficient to determine the functions $F_{j,\pm j}$ completely.
As shown in detail in appendix \ref{sec:AppellF4}, the system of partial differential equation generated by $\A^\one_{\Delta_1, \Delta_2; \Delta, m}$ and $\A^\two_{\Delta_1, \Delta_2}$ is of the Appell $F_4$ type.
It admits as a solution the double hypergeometric series
\begin{equation}
	F_4\left( \tfrac{6 - \Delta_1 - \Delta_2 - \Delta + m}{2}, 
	\tfrac{3 - \Delta_1 - \Delta_2 + \Delta + m}{2},
	\tfrac{5}{2} - \Delta_1, \tfrac{5}{2} - \Delta_2 ;
	x(1-y), (1-x)y \right),
	\label{eq:F4}
\end{equation}
defined in eq.~\eqref{eq:F4:def}.
Note that the arguments of this function are precisely $p_1^2/(p_1+p_2)^2$ and $p_2^2/(p_1 + p_2)^2$, as defined by eq.~\eqref{eq:xy}.
This is the unique solution that is analytic around $(x,y) = (0,0)$ (up to an overall multiplicative constant), but there are in fact other solutions that cannot be expressed as Taylor series in $x$ and $y$. 
For generic values of the scaling dimensions, the most general solution to the system \eqref{eq:Fjj:two}--\eqref{eq:Fjj:one} is a linear combination of the following four functions~\cite{Exton:1995}:
\begin{equation}
\begin{aligned}
	& w_1^{\Delta_1 - 3/2} w_2^{\Delta_2 - 3/2}
	f_{\Delta_1,\Delta_2, \Delta, j, j}(x,y),
	\\
	& w_1^{\Delta_1 - 3/2}
	f_{\Delta_1, 3 - \Delta_2, \Delta, j, j}(x,y),
	\\
	& w_2^{\Delta_2 - 3/2}
	f_{3-\Delta_1, \Delta_2, \Delta, j, j}(x,y),
	\\
	& f_{3-\Delta_1, 3- \Delta_2, \Delta, j, j}(x,y),
\end{aligned}
\label{eq:4solutions}
\end{equation}
where we have used the shorthand notation $w_1 = x (1-y)$ and $w_2 = y(1-x)$, corresponding to the ratios of momenta squared, and defined the function
\begin{align}
	f_{\Delta_1, \Delta_2, \Delta, j, j}(x,y) &= (1 - x - y)^j
	\nonumber \\
	& \quad \times
	F_4\left( \tfrac{\Delta_1 + \Delta_2 - \Delta + j}{2},
	\tfrac{\Delta_1 + \Delta_2 + \Delta - 3 + j}{2},
	\Delta_1 - \tfrac{1}{2}, \Delta_2 - \tfrac{1}{2};
	x(1-y), (1-x)y \right).
	\label{eq:fjj}
\end{align}
The fact that the conformal Ward identities are solved by Appell $F_4$ functions in terms of the ratios of momenta squared was first observed in the scalar case $j = 0$~\cite{Coriano:2013jba, Bzowski:2013sza}.
Here we have just shown that the highest-helicity components of the 3-point function must equally be of the Appell $F_4$ type, up to a multiplicative factor of $(1-x-y)^j$. This fact was already observed in refs.~\cite{Gillioz:2019lgs, Gillioz:2020wgw}.

The advantage of working in the helicity basis is that the recursion relation can be solved exactly. The result is not particularly pleasing for the eye, but it is a function of the generalized hypergeometric type that can be evaluated efficiently at any value of $x$ and $y$: as explained in appendix~\ref{sec:AppellF4}, we find
\begin{equation}
	f_{\Delta_1, \Delta_2, \Delta, j, m}(x,y)
	= \sum_{n = 0}^{j - m} (1-x-y)^{j-n}
	\sum_{i = 0}^n
	\chi_{\Delta_1, \Delta_2, \Delta, j, m}^{(n,i)}
	f_{\Delta_1, \Delta_2, \Delta, j}^{(n-i, i)}(x,y)
	\label{eq:fjm}
\end{equation}
where $f^{(a,b)}$ is a generalization of the Appell $F_4$ series, defined by the infinite series
\begin{align}
	f_{\Delta_1, \Delta_2, \Delta, j}^{(a,b)}(x,y)
	= \sum_{r = 0}^\infty &
	\frac{\left( \frac{\Delta_1 + \Delta_2 - \Delta + j}{2} \right)_r
	\left( \frac{\Delta_1 + \Delta_2 + \Delta - 3 + j}{2} \right)_r
	\left( j + \frac{1}{2} \right)_r}
	{r! \left( \Delta_1 - \frac{1}{2} + a \right)_r
	\left( \Delta_2 - \frac{1}{2} + b \right)_r}
	x^{a + r} y^{b + r}
	\nonumber \\
	& \times
	{}_2F_1\left( \tfrac{\Delta_1 + \Delta_2 - \Delta + j}{2} + r,
	\tfrac{\Delta_1 + \Delta_2 + \Delta - 3 + j}{2} + r,
	\Delta_1 - \tfrac{1}{2} + a + r; x \right)
	\nonumber \\
	& \times
	{}_2F_1\left( \tfrac{\Delta_1 + \Delta_2 - \Delta + j}{2} + r,
	\tfrac{\Delta_1 + \Delta_2 + \Delta - 3 + j}{2} + r,
	\Delta_2 - \tfrac{1}{2} + b + r; y \right),
	\label{eq:fab}
\end{align}
and the coefficients of the linear combination are
\begin{align}
	\chi_{\Delta_1, \Delta_2, \Delta, j, m}^{(n,i)}
	&= \frac{(-1)^{j + m + n + i}}{i! (n - i)!}
	\frac{4^{j - m} (m + 1)_{j - m}}
	{(2m + 1)_{j - m - n}}
	\frac{\left( \frac{\Delta + j - \Delta_{12}}{2} - n + i \right)_{n-i}
	\left( \frac{\Delta + j + \Delta_{12}}{2} - i \right)_i}
	{\left( \Delta_1 - \frac{1}{2} \right)_{n-i}
	\left( \Delta_2 - \frac{1}{2} \right)_i
	(\Delta + m - 1)_{j - m}}
	\nonumber \\
	\times  &
	\sum_{k = 0}^{j - m - n}
	\frac{(-1)^k
	\left( m + \frac{1}{2} \right)_{j - m - n - k}
	\left( m + \frac{1}{2} \right)_k}
	{k! (j - m - n - k)!}
	\left( \tfrac{\Delta - j - 1 + \Delta_{12}}{2} \right)_{j - m - i - k}
	\left( \tfrac{\Delta - j - 1 - \Delta_{12}}{2} \right)_{i + k},
	\label{eq:chi}
\end{align}
where we have denoted $\Delta_{12} = \Delta_1 - \Delta_2$.
This is the result of applying the recursion relation \eqref{eq:Fjm:ladder} starting from the highest helicity solution given in eq.~\eqref{eq:fjj}. The other solutions can be obtained from this one applying the ``shadow transform'' $\Delta_i \leftrightarrow 3 - \Delta_i$ to any of the two scaling dimensions $\Delta_1$ and $\Delta_2$, and multiplying with the appropriate power of $w_1$ and $w_2$ as in eq.~\eqref{eq:4solutions}.

Some important properties of the 3-point function can be deduced from this general solution:
\begin{itemize}

\item[$\circ$]
If $j$ is half-integer, the series generated by the recursion relation never terminates, which is inconsistent with our ansatz \eqref{eq:3pt:helicity}. It is of course well known that only integer spin operators enter the OPE of two scalars, so that this pathological situation is never realized. With integer $j$, one can verify that once the recursion relation passes $m = 0$, the linear combination involves fewer terms again until it finally collapses back to a single term at $m = -j$.
In fact, the symmetry of eq.~\eqref{eq:Fjm:ladder} around $m = 0$ implies that
\begin{equation}
	f_{\Delta_1, \Delta_2, \Delta, j, -m}(x,y)
	= f_{\Delta_1, \Delta_2, \Delta, j, m}(x,y).
	\label{eq:helicity:symmetry}
\end{equation}
Therefore, all functions can be expressed in terms of eq.~\eqref{eq:fjm}
with $m \geq 0$.

\item[$\circ$]
The formula \eqref{eq:fjm} involves terms with every positive power of $(1 - x - y)$ between $m$ and $j$. In the \emph{colinear limit} $p_1 \propto p_2$, we have $1 - x - y = 0$ and therefore all functions $f$ vanish identically except for the one with $m = 0$.
This is because there is no angular momentum in this case, and therefore the vanishing spin of the state created by the two scalar operators $\phi_1$ and $\phi_2$ must be matched by the spin $m$ of the state created by the operator $\O$.
Eq.~\eqref{eq:fjm:colinear} in the appendix gives the simpler form of the solution in that limit.

\item[$\circ$]
Since the differential operator $\A^\two_{\Delta_1, \Delta_2}$ involved in the recursion relation \eqref{eq:Fjm:ladder} is odd under the simultaneous exchange $\Delta_1 \leftrightarrow \Delta_2$ and $x \leftrightarrow y$, the solution must satisfy
\begin{equation}
	f_{\Delta_2, \Delta_1, \Delta, j, m}(y,x)
	= (-1)^{j-m} f_{\Delta_1, \Delta_2, \Delta, j, m}(x,y)
\end{equation}
This is indeed what is observed: the generalization \eqref{eq:fab} of the Appell $F_4$ function has this symmetry upon exchanging $a \leftrightarrow b$, and similarly the coefficients $\chi$ transform into each other under $i \leftrightarrow n - i$.
A consequence of this is that when $\Delta_1 = \Delta_2$, for instance when the two scalar operators are identical, then the operator $\O$ can only have  even spin. This is again a well-known property of the OPE of two identical scalar operators.

\item[$\circ$]
Finally, it should be noted that all coefficients $\chi$ with $m < j$ vanish if $\Delta_1 = \Delta_2$ and $\Delta = j + 1$. This corresponds to the case in which $\O$ is a conserved current, transforming in a short representation of the conformal group. The vanishing of the Wightman 3-point function is consistent with our findings of section~\ref{sec:2pt} in which we observed that $\O$ only creates states of helicity $m = \pm j$.

\end{itemize}

\subsection{Correlation function}

We are now armed with a set of functions that satisfy all constraints imposed by conformal symmetry. But what linear combination of solutions corresponds to the Wightman correlation function?

Conformal Ward identities alone are not enough to answer this question. In fact, for generic (non-integer) value of the scaling dimensions, different correlation functions obey the same set of identities: Wightman functions, time-ordered products, and retarded products can all be expressed as linear combinations of the solutions that we just found.
Additional information is needed to distinguish between these various cases. It can be obtained using analyticity properties that follow from the micro-causality condition~\cite{Gillioz:2021sce}, or in the specific case of Wightman functions, studying some OPE limits~\cite{Gillioz:2019lgs}. 
We refer the reader to the relevant work for more details, and focus here on the result.

In the case where both $p_1$ and $p_2$ are inside the forward light cone, one finds
\begin{align}
	F_{j,m}(x, y) &= 4 \pi \N \,
	\frac{\sin\left[ \pi
	\frac{\Delta_1 - \Delta_2 - \Delta + j + 3}{2} \right]}
	{\Gamma\left( \Delta_1 - \frac{1}{2} \right)}
	w_1^{\Delta_1 - 3/2}
	\nonumber \\
	& \quad \times \bigg[
	\Gamma\left( \tfrac{3}{2} - \Delta_2 \right)
	\Gamma\left( \tfrac{\Delta_1 + \Delta_2 - \Delta + j}{2} \right)
	\Gamma\left( \tfrac{\Delta_1 + \Delta_2 + \Delta + j - 3}{2} \right)
	w_2^{\Delta_2 - 3/2}
	f_{\Delta_1, \Delta_2, \Delta, j, m}(x, y)
	\nonumber \\
	& \quad\qquad 
	+ \Gamma\left( \Delta_2 - \tfrac{3}{2} \right)
	\Gamma\left( \tfrac{\Delta_1 - \Delta_2 - \Delta + j + 3}{2} \right)
	\Gamma\left( \tfrac{\Delta_1 - \Delta_2 + \Delta + j}{2} \right)
	f_{\Delta_1, 3 - \Delta_2, \Delta, j, m}(x, y) \bigg].
	\label{eq:Fjm}
\end{align}
This is the unique linear combination of solutions that is on the one hand proportional to $w_1^{\Delta_1 - 3/2}$ as $p_1^2 \to 0$ (the other two solutions are analytic in $w_1$), in accordance with the normalization of the state $\phi_1(p_1) \ket{0}$ found in section~\ref{sec:2pt}, and on the other hand that is also analytic in $1 - w_2$ as $w_2 \to 1_-$, as required by the existence of the limit $p_1^\mu \to 0$.
The constant $\N$ is related to the OPE coefficient, and it is therefore arbitrary unless we adopt some specific convention. Following one standard choice detailed in ref.~\cite{Gillioz:2020wgw} in which the OPE coefficient  $\lambda_{12\O}$ is defined from the position-space correlation function, we have
\begin{equation}
	\N = \frac{(4\pi)^d (-i)^j (\Delta - 1)_j \, \lambda_{12\O}}
	{2^{\Delta_1 + \Delta_2 + \Delta}
	\Gamma\left( \frac{\Delta_1 - \Delta_2 + \Delta + j}{2} \right)
	\Gamma\left( \frac{\Delta_2 - \Delta_1 + \Delta + j}{2} \right)
	\Gamma\left( \frac{\Delta_1 + \Delta_2 - \Delta + j}{2} \right)
	\Gamma\left( \frac{\Delta_1 + \Delta_2 + \Delta + j - 3}{2} \right)}.
\end{equation}

When $p_2$ is instead space-like, we have
\begin{align}
	F_{j,m}(x, y) &= 4 \pi^2 \N \,
	\frac{1}{\Gamma\left( \Delta_1 - \frac{1}{2} \right)}
	w_1^{\Delta_1 - 3/2}
	\nonumber \\
	& \quad \times \bigg[
	\frac{\Gamma\left( \frac{3}{2} - \Delta_2 \right)
	\Gamma\left( \frac{\Delta_1 + \Delta_2 + \Delta + j - 3}{2} \right)}
	{\Gamma\left( 1 - \frac{\Delta_1 + \Delta_2 - \Delta + j}{2} \right)}
	(-w_2)^{\Delta_2 - 3/2}
	f_{\Delta_1, \Delta_2, \Delta, j, m}(x, y)
	\nonumber \\
	& \quad\qquad 
	+ \frac{\Gamma\left( \Delta_2 - \frac{3}{2} \right)
	\Gamma\left( \frac{\Delta_1 - \Delta_2 + \Delta + j}{2} \right)}
	{\Gamma\left( 1- \frac{\Delta_1 - \Delta_2 - \Delta + j + 3}{2} \right)}
	f_{\Delta_1, 3 - \Delta_2, \Delta, j, m}(x, y) \bigg].
	\label{eq:Fjm:spacelike}
\end{align}
This only differs from eq.~\eqref{eq:Fjm} by a ratio of phases in the first term in square brackets. In this case, the highest-helicity function with $m = +j$ takes a particularly simple form thanks to a property of the Appell $F_4$ function: it can be written
\begin{align}
	F_{j,j}(x, y) &= 4 \pi^2 \N \,
	\frac{\Gamma\left( \frac{\Delta_1 - \Delta_2 + \Delta + j}{2} \right)
	\Gamma\left( \frac{\Delta_1 + \Delta_2 + \Delta + j - 3}{2} \right)}
	{\Gamma\left( \Delta_1 - \frac{1}{2} \right)
	\Gamma\left( \Delta - \frac{1}{2} \right)}	
	\nonumber \\
	& \quad \times w_1^{\Delta_1 - 3/2}
	(-w_2)^{-(\Delta_1 - \Delta_2 + \Delta + j)/2}
	f_{\Delta_1, \Delta, \Delta_2, j, j}
	\left( \frac{x}{x-1}, \frac{1}{y} \right),
	\label{eq:Fjj:spacelike}
\end{align}
in terms of a unique Appell $F_4$ hypergeometric function in which the roles of $\Delta_2$ and $\Delta$ have been exchanged. In this case the correlation function is still proportional to $w_1^{\Delta_1 - 3/2}$ as $p_1^2 \to 0$, but it has also the scaling limit of a state created by the operator $\O$ when $(p_1 + p_2)^2 \to 0$.
Unfortunately, the type of transformation property relating series expansions around $(x, y) = (0,0)$ and $(x, y) = (0, -\infty)$ is special to the Appell $F_4$ function and does not carry on in a simple manner to its generalization \eqref{eq:fab}. For this reason, the problem of finding a closed form expression for the function $F_{j, m}$ with $m \neq \pm j$ when $p_2$ is outside the forward light cone will be treated elsewhere.%
\footnote{In fact, solving the conformal Ward identities in the case where all 3 operators carry spin will contribute to answering this question, as one can then always choose the reference momentum --- the one expressed in terms of spinors as in eq.~\eqref{eq:p} --- to be the largest so that $x$ and $y$ remain small.
Our approach in this work relies on the contrary on taking the momentum of the \emph{spinning} operator as a reference, and therefore it is not general enough to obtain a simple solution.}
Note that eq.~\eqref{eq:Fjm:spacelike} gives a partial answer, but it involves the function \eqref{eq:fab} whose definition as a series expansion is only convenient in a neighborhood of $(x,y) = (0,0)$.
Nevertheless, even in the absence of closed-form solution, the recursion relation \eqref{eq:Fjm:ladder} can still be used directly onto eq.~\eqref{eq:Fjj:spacelike}.

The results presented in this section are the main output of this work. We have shown how the parameterization of the 3-point function with spinors helps express the conformal Ward identities in terms of simple ladder operators, how the various helicity eigenstates are related by the action of a differential operator, and we even arrived at a closed-form solution valid when $p_1^2$ and $p_2^2$ are small compared to $(p_1 + p_2)^2$.
Before concluding, we want to illustrate in the next section how this 3-point function can be used as elementary building block for the construction of higher-point Wightman functions.


\section{The OPE for higher-point functions}
\label{sec:4pt}

Our results for the 2- and 3-point correlation functions implicitly define a momentum-space OPE for conformal primary operators acting on the vacuum, and the goal of this short section is to illustrate its conceptual simplicity.

The existence of an operator product expansion in quantum field theory is intrinsically related to the completeness of the Hilbert space: if all the states created by a single local operator acting on the vacuum span the entire Hilbert space, then the product of two (or more) operators must also be expressible in that basis. In other words, we must be able to write
\begin{equation}
	\phi_2(p_2) \phi_1(p_1) \ket{0} 
	\approx
	\sum_\O C_{12\O} \O(p_1 + p_2) \ket{0},
	\label{eq:OPE:sketch}
\end{equation}
where $C_{12\O}$ is an object that depends on the momenta $p_1$ and $p_2$.
This statement can be made sharp in conformal field theory thanks to radial quantization in Euclidean position space.%
\footnote{Another salient feature of radial quantization is that the convergence rate of the OPE is controlled by the separation of operators in space, and this can be used to show that the OPE is absolutely convergent over most of the configuration space. There is no such argument in momentum space, and therefore we generically only expect convergence in the sense of distributions (see however ref.~\cite{Gillioz:2019iye} for a counterexample in 2 dimensions).}
In this case however the sum does not only involve primary operators on the right-hand side, but also descendants. 
This inconvenience disappears when working in momentum space, as there is a unique momentum eigenstate in each conformal family that can appear on the right-hand side of eq.~\eqref{eq:OPE:sketch}.

The other difficulty in defining the OPE has to do with spin: even when the operators appearing on the left-hand side of eq.~\eqref{eq:OPE:sketch} are scalars, the fact that they are inserted at separated points --- or come with distinct momenta --- means that the state carries angular momentum, and thus that the operator on the right-hand side might have spin.
This is where the spinor-helicity formalism becomes extremely useful, as it allows to further decompose the OPE \eqref{eq:OPE:sketch} into helicity eigenstates.%
\footnote{Strictly speaking the spinor-helicity formalism is not necessary to do so: when the two operators are scalars, the total spin is given by the angular momentum, which means that the polarization tensors can be constructed out of the momenta $p_1$ and $p_2$ only. This fact has been used in ref.~\cite{Gillioz:2020wgw} to write down the OPE for a scalar 4-point function in general space-time dimension without making use of spinors.}
The decomposition~\eqref{eq:helicityeigenstates} can be used to absorb the polarization tensors into the object $C_{12\O}$ and write
\begin{equation}
	\phi_2(p_2) \phi_1(p_1) \ket{0} 
	= \sum_\O \sum_{m = -j}^j
	 C^{(j,m)}_{12\O} \O_{j,m}(p_1 + p_2) \ket{0}.
	\label{eq:OPE}
\end{equation}
$C_{12\O}$ is now a mere function of the momenta $p_1$ and $p_2$, which can be computed taking the overlap of both sides of this equation with a helicity eigenstate $\O_{j,m}(p) \ket{0}$.
We find
\begin{equation}
	C^{(j,m)}_{12\O} = 
	\frac{\llangle \O_{j,m}(-p) \phi_2(p_2) \phi_1(p_1) \rrangle}
	{\llangle \O_{j,m}(-p) \O_{j,m}(p) \rrangle}
	= (p^2)^{(\Delta_1 + \Delta_2 - \Delta + j - 3)/2}
	C_{\Delta, j, m}^{-1}
	F_{j,m}(x, y),
	\label{eq:OPE:coefficient}
\end{equation}
where the coefficient $C_{\Delta, j, m}$ is given in eq.~\eqref{eq:C}, the function $F_{j,m}$ in eq.~\eqref{eq:Fjm}, and $x$, $y$ are defined in terms of the momenta $p_1$ and $p_2$ in eq.~\eqref{eq:xy}.
The dynamical OPE coefficient is contained in the definition of $F_{j,m}$.

As it is, this result is valid when the helicity of the operator $\O_{j,m}$ on the right-hand side of eq.~\eqref{eq:OPE} is defined with respect to the same spinors $\lambda$ and $\lambdabar$ used to parameterized the momenta $p_1$ and $p_2$ as in eq.~\eqref{eq:p1:p2:lambda}.
This is not necessary, however, and in fact it is not wanted in the perspective of iterating the OPE.
Thankfully, the freedom to choose a different basis of spinors is easy to implement in eq.~\eqref{eq:OPE}: 
any other basis will be related to $\{ \lambda^a, \lambdabar^a \}$ by a little group transformation~\eqref{eq:littlegroup}. If we define this new basis by
\begin{equation}
	\lambda'^a = e^{i\theta} \lambda^a,
	\qquad\qquad
	\lambdabar'^a = e^{-i\theta} \lambdabar^a,
\end{equation}
then we can again compute the overlap of eq.~\eqref{eq:OPE} with an eigenstate of helicity given in the original basis.
This time, the computation involves contracting the indices of two polarization tensors of the form~\eqref{eq:polarization} that satisfy
\begin{equation}
	\varepsilon'_{-m'} \cdot \varepsilon_m = \delta_{m'm} e^{i \theta m}
	(-1)^j (p^2)^j.
\end{equation}
In other words, the change of basis only accounts for a phase $e^{i \theta m}$ in the coefficient \eqref{eq:OPE:coefficient}, which is trivial to implement in a computation.

To give a concrete example, this OPE can be used to decompose the 4-point correlation function of scalar primary operators into conformal partial waves. We find immediately
\begin{equation}
	\llangle \phi_4(p_4) \phi_3(p_3) \phi_2(p_2) \phi_1(p_1) \rrangle
	= s^{(\Delta_\Sigma - 9)/2} 
	\sum_\O \sum_{m = -j}^j
	C_{\Delta, j, m}^{-1}
	e^{i m \theta}
	F_{j,-m}(x',y')^*
	F_{j,m}(x,y)
\end{equation}
where $s = (p_1 + p_2)^2 = (p_3 + p_4)^2$ is the square of the center-of-mass energy, the variables $x$, $y$, $x'$ and $y'$ are defined by
\begin{equation}
	x (1-y) = \frac{p_1^2}{s},
	\qquad
	(1-x) y = \frac{p_2^2}{s},
	\qquad
	(1-x') y' = \frac{p_3^2}{s},
	\qquad
	x' (1-y') = \frac{p_4^2}{s},
\end{equation}
and $\theta$ is the angle formed by the space-like vector $q$ defined in terms of momenta $p_1$ and $p_2$ in eq.~\eqref{eq:q:def} and a similar vector defined in terms of $p_3$ and $p_4$. We have also denoted $\Delta_\Sigma = \Delta_1 + \Delta_2 + \Delta_3 + \Delta_4$.
Note that because of the symmetry relation \eqref{eq:helicity:symmetry} the phase always appears in the linear combination $e^{i m \theta} + e^{-i m \theta} = 2 \cos(m \theta)$, which is (twice) the Chebyshev polynomial $T_m(\cos\theta)$ corresponding to spherical harmonics in 3 space-time dimensions. This result precisely matches the conformal partial waves obtained in ref.~\cite{Gillioz:2020wgw} in the special case $d = 3$.
A significant improvement with respect to that work is that the 3-point functions are now known exactly, whereas they were given by a recursion relation there.%
\footnote{The two-step recursion relation for the 3-point function given in ref.~\cite{Gillioz:2020wgw} is a linear combination of the two equations corresponding to both signs in eq.~\eqref{eq:Fjm:ladder}.
}

\section{Conclusions}
\label{sec:conclusions}

This work develops a new formalism for conformal correlators in terms of spinor variables.
We have emphasized the usefulness of spinors to decompose the states created by a single conformal primary operator acting on the vacuum into helicity eigenstates, and extended this formalism to the product of two operators acting on the vacuum.
While the 3-point functions involved in these computation are not completely new, the novel formalism that we used makes their derivation particularly clear when compared with other methods.

This work is also an important milestone in a larger research program: the construction of all conformal Wightman functions in Minkowski momentum space. The state-of-the-art in this area can be summarized as follows:
\begin{enumerate}

\item
Wightman 2-point functions in momentum space have been known since the early days of conformal field theory, and used in the constructions of all unitary representations~\cite{Mack:1975je}.
They recently re-appeared as an efficient way of computing Euclidean conformal blocks in ref.~\cite{Erramilli:2019njx}.
For tensor operators, a construction valid in any space-time dimension has been presented in ref.~\cite{Gillioz:2018mto}.

\item
The Wightman 3-point function of scalar operators was constructed for the first time in closed-form in ref.~\cite{Gillioz:2019lgs}.
An integral representation particularly useful for holographic applications is also available~\cite{Bautista:2019qxj}.

\item
The 3-point function of two scalars and one symmetric tensor operator was also presented in ref.~\cite{Gillioz:2019lgs}, but its projection onto spin eigenstates only appeared in ref.~\cite{Gillioz:2020wgw}. This allowed for a convenient decomposition of the 4-point function into partial waves in the same work.%
\footnote{An interesting feature of all these results is that they are analytic in the space-time dimension $d$. This is a peculiarity of working only with tensor representations of the Lorentz group that exist in any $d$.}

\end{enumerate}
The obvious next step is to compute Wightman 3-point functions of operators carrying arbitrary Lorentz representations. This would not only grant access to partial wave expansions for generic 4-point functions, but also for any higher-point function through the momentum-space OPE.%
\footnote{The current formalism only allows for the computation of higher-point partial waves in which one of the intermediate operator is a scalar.} 

The present work does not yet present results for these generic 3-point function, but it takes an important step in this direction. The derivation of conformal Ward identities in terms of helicity ladder operators in not conceptually different from what has been presented here. It is however vastly more complicated to solve the resulting partial differential equations, and for this reason the results will be presented in a separated work.
The goal of the present paper is therefore to introduce the method, and illustrate its simplicity when applied to 3-point correlation functions involving two scalar operators.
For the same reason, we have chosen to work here exclusively in 3 space-time dimensions. The results will be generalized to conformal field theory in 4 and higher dimensions elsewhere.

The informed reader will also wonder what is the physical purpose of this program, besides its mathematical interest. There is in fact already a large body of literature on the construction of correlation functions that have the isometries of the Euclidean conformal group $\SO(d+1, 1)$, 
both for CFT in flat space, of for QFT in AdS. Many works have been specifically dedicated to the study of operators transforming in various non-trivial representations~\cite{Costa:2014kfa, Costa:2018mcg, Meltzer:2019nbs, Fortin:2019dnq}.
In fact, the problem has essentially been solved by the introduction of weight-shifting operators~\cite{Sleight:2017fpc, Isono:2017grm, Karateev:2017jgd, Karateev:2018oml}.
One aspect in which our approach differs from the standard literature is that it relies on the Lorentzian conformal group $\SO(d, 2)$. However, one might argue that the two are substantially equivalent thanks to the simple analytic continuation that is possible between Euclidean correlators and Wightman functions in Minkowski space-time~\cite{Kravchuk:2021kwe}.

Nevertheless, there are at least three directions of research that will actively benefit from the development of this program.
The first one is the study of higher-point correlation functions, both in CFT and in AdS. The computation of conformal blocks for functions of more than 4 points is quite difficult, and various methods have been used~\cite{Rosenhaus:2018zqn, Goncalves:2019znr, Fortin:2019zkm, Bercini:2020msp, Poland:2021xjs, Buric:2021ywo, Buric:2021kgy, Antunes:2021kmm}.
It is obvious that the factorization observed in the Minkowski momentum space OPE present important advantages, even though it only applies to Wightman functions.
The second direction of research is Hamiltonian truncation in infinite volume~\cite{Katz:2016hxp, Anand:2020gnn}, where one of the computational bottlenecks is the evaluation of matrix elements that correspond to a particular limit of the Wightman 3-point functions~\cite{Anand:2019lkt}.
Finally, the last area in which this work can have an impact is quantum field theory in de~Sitter space-time. Many recent works have exploited the Euclidean conformal symmetry of late-time correlators~\cite{Arkani-Hamed:2018kmz, Baumann:2019oyu, Baumann:2020dch, Hogervorst:2021uvp, DiPietro:2021sjt, Pethybridge:2021rwf}, but the Lorentzian conformal group also plays an important role in the in-in formalism recently put forward~\cite{Higuchi:2010xt, Gorbenko:2019rza, Sleight:2020obc, Sleight:2021plv}.


\subsection*{Acknowledgments}

The author would like to thank Brian Henning and Matthew Walters for stimulating discussions, as well as the University of Z\"urich for their hospitality during the completion of this work.


\appendix
\section{Conventions for spinors}
\label{sec:conventions}

We work in 3-dimensional Minkowski space-time with the mostly-minus metric%
\footnote{While the mostly-plus metric is arguably the best choice in position space, as it makes the analytic continuation between Euclidean and Minkowski spaces obvious, the mostly-minus metric is definitely more convenient to work with in momentum space.}
$\eta_{\mu\nu} = \diag(1, -1, -1)_{\mu\nu}$, where $\mu = 0, 1, 2$.
Let $z^\mu$ be a null vector pointing in the forward direction,
\begin{equation}
	z^2 = (z^0)^2 - (z^1)^2 - (z^2)^2 = 0,
	\qquad\qquad
	z^0 > 0.
\end{equation} 
$z^\mu$ is conveniently parameterized in terms of a two-component real spinor $\zeta^a$ as
\begin{equation}
	z^\mu = \tfrac{1}{2} \, \sigma^\mu_{ab} \zeta^a \zeta^b,
	\label{eq:z:spinors}
\end{equation}
where repeated indices $a,b = 1,2$ are summed over, and $\sigma^\mu$ are the $2 \times 2$ symmetric matrices
\begin{equation}
	\sigma^0_{ab} = \left(\begin{array}{cc}
		1 & 0 \\ 0 & 1
	\end{array}\right)_{ab},
	\qquad
	\sigma^1_{ab} = \left(\begin{array}{cc}
		0 & 1 \\ 1 & 0
	\end{array}\right)_{ab},
	\qquad
	\sigma^2_{ab} = \left(\begin{array}{cc}
		1 & 0 \\ 0 & -1
	\end{array}\right)_{ab}.
\end{equation}
We may sometimes also use the 2-index notation for a vector, $z^\mu \equiv \tfrac{1}{2} \, \sigma^\mu_{ab} z^{ab}$.
Introducing 
\begin{equation}
	\varepsilon_{ab} = \left(\begin{array}{cc}
		0 & 1 \\ -1 & 0
	\end{array}\right)_{ab},
\end{equation}
we have the identity
\begin{equation}
	\eta_{\mu\nu} \sigma^\mu_{ab} \sigma^\nu_{cd}
	= \varepsilon_{ac} \varepsilon_{bd}
	+ \varepsilon_{ad} \varepsilon_{bc}.
\end{equation}
Lorentz-invariant quantities can therefore be constructed out of spinors with upper indices contracted with the matrix $\varepsilon$ with lower indices: for instance, the antisymmetric scalar product of two spinors is defined by
\begin{equation}
	\zeta \cdot \xi \equiv \varepsilon_{ab} \zeta^a \xi^b
	= - \xi \cdot \zeta.
\end{equation} 
For $z^\mu$ defined in eq.~\eqref{eq:z:spinors} and another vector $x^\mu = \tfrac{1}{2} \, \sigma^\mu_{ab} \xi^a \xi^b$, we have 
\begin{equation}
	z \cdot x
	= \eta_{\mu\nu} z^\mu x^\nu
	= \tfrac{1}{4} \, \eta_{\mu\nu} \sigma^\mu_{ab} \sigma^\nu_{cd}
	\zeta^a \zeta^b \xi^c \xi^d	
	= \tfrac{1}{2} \, \varepsilon_{ac} \varepsilon_{bd}
	\zeta^a \zeta^b \xi^c \xi^d	
	= \frac{1}{2} (\zeta \cdot x)^2.
\end{equation}
It is sometimes convenient to introduce the contravariant spinor
\begin{equation}
	\zeta_a \equiv \varepsilon_{ab} \zeta^b,
\end{equation}
in terms of which the scalar product can be written
\begin{equation}
	\zeta \cdot \xi = \zeta^a \xi_a = - \zeta_a \xi^a.
\end{equation}
A lower spinor index can be raised with the inverse tensor $\varepsilon^{ab} = -\varepsilon^{ba}$ satisfying
\begin{equation}
	\varepsilon^{ac} \varepsilon_{cb}
	= \varepsilon_{bc} \varepsilon^{ca} = \delta^a_b,
\end{equation}
so that $\varepsilon^{ab} \zeta_b = \zeta^a$. 
Note that both the (inverse) metric $\eta^{\mu\nu}$ and the Levi-Civita symbol $\varepsilon^{\mu\nu\rho}$ can be expressed in terms of $\sigma^\mu$ through the identity
\begin{equation}
	(\sigma^\mu \cdot \sigma^\nu)_{ab}
	\equiv
	- \sigma^\mu_{ac} \varepsilon^{cd} \sigma^\nu_{db} 
	= \eta^{\mu\nu} \varepsilon_{ab}
	+ \varepsilon^{\mu\nu\rho} (\sigma_\rho)_{ab}.
	\label{eq:sigma:product}
\end{equation}

\paragraph{Complex spinors.}

Since spinors have two components, any pair of linearly independent spinors can be used as a basis. We shall take the real spinors $\lambda_1^a$ and $\lambda_2^a$ to form such a basis, or more conveniently the complex spinors $\lambda^a$ and $\lambdabar^a$ that are conjugate to each other, defined by
\begin{equation}
	\lambda^a = \frac{\lambda_1^a + i \lambda_2^a}{\sqrt{2}},
	\qquad\qquad
	\lambdabar^a = \frac{\lambda_1^a - i \lambda_2^a}{\sqrt{2}}.
\end{equation}
This means that we can write
\begin{equation}
	\varepsilon^{ab} =
	\frac{\lambdabar^a \lambda^b - \lambda^a \lambdabar^b}
	{\lambda \cdot \lambdabar},
	\label{eq:epsilon:lambda}
\end{equation}
or express any other spinor in this basis through
\begin{equation}
	\zeta^a = 
	\frac{(\zeta \cdot \lambdabar)\lambda^a
	- (\zeta \cdot \lambda) \lambdabar^a }
	{\lambda \cdot \lambdabar}.
\end{equation}
These two spinors can be used to parameterize a momentum inside the forward light cone as in eq.~\eqref{eq:p},
\begin{equation}
	p^\mu = \sigma^\mu_{ab} \, \lambda^a \lambdabar^b
	= \tfrac{1}{2} \sigma^\mu_{ab} \left( \lambda_1^a \lambda_1^b
	+ \lambda_2^a \lambda_2^b \right).
\end{equation}
Note that the energy $p^0 = \left| \lambda_1 \right|^2 + \left| \lambda_2 \right|^2$ is positive, and likewise
\begin{equation}
	p^2 = - (\lambda \cdot \lambdabar)^2
	= (\lambda_1 \cdot \lambda_2)^2 > 0.
\end{equation}
With our conventions the scalar product $\lambda \cdot \lambdabar$ is purely imaginary.

\paragraph{Spinor derivatives.}

The spinor derivative is defined by the equation
\begin{equation}
	\frac{\partial \zeta^b}{\partial \zeta^a} = \delta_a^b,
\end{equation}
or equivalently
\begin{equation}
	\frac{\partial \zeta^b}{\partial \zeta_a}
	= \varepsilon^{ab},
\end{equation}
which implies for the scalar product
\begin{equation}
	\frac{\partial}{\partial \zeta^a} (\zeta \cdot \xi)
	= \xi_a,
	\qquad
	\frac{\partial}{\partial \xi^a} (\zeta \cdot \xi)
	= - \zeta_a.
\end{equation}
When applied to a function of $z^\mu$, the chain rule gives
\begin{equation}
	\frac{\partial f(z)}{\partial \zeta^a}
	= \sigma^\mu_{ab} \zeta^b \frac{\partial f(z)}{\partial z^\mu}.
\end{equation}
Using the identity \eqref{eq:sigma:product}, we can form the combination
\begin{equation}
	\sigma^\mu_{ab} \zeta^a \frac{\partial f(z)}{\partial \zeta_b}
	= \varepsilon^{\mu\nu\rho} z_\nu
	\frac{\partial f(z)}{\partial z^\rho}.
\end{equation}
This is precisely the derivative operator appearing in the definition \eqref{eq:spin} of the spin operator: when applied to a correlation function in which Lorentz indices are contracted with an auxiliary vector $z^\mu$, the spin operator takes the familiar form~\cite{Costa:2011mg, Dobrev:1975ru}
\begin{equation}
	\widehat{\Sigma}_{\mu\nu}(z)
	= z_\mu \frac{\partial}{\partial z^\nu}
	- z_\nu \frac{\partial}{\partial z^\mu}.
\end{equation}

\paragraph{Derivatives with respect to momenta.}

Expressing derivatives with respect to momenta in terms of spinors is the key element in the computation of the conformal Ward identities for 2- and 3-point functions. In the presence of a single momentum $p$, the form of this derivative was given in eq.~\eqref{eq:dp}, and its projection onto helicity components in eq.~\eqref{eq:dp:projection}.
This led to the representations \eqref{eq:D} and \eqref{eq:M} for the generators of scale and Lorentz transformations. For special conformal transformations, the operator $\K^\mu$ defined in eq.~\eqref{eq:K:2pt:p} is given by
\begin{align}
	\K^\mu = \tfrac{1}{2} \sigma^\mu_{ab} \bigg[ \, &
	-\frac{\partial^2}{\partial \lambda_a \partial \lambdabar_b}
	+ \frac{\Delta - 2}{\lambda \cdot \lambdabar}
	\left( \lambdabar^a \frac{\partial}{\partial \lambdabar_b}
	- \lambda^a \frac{\partial}{\partial \lambda_b} \right)
	\nonumber \\
	& + \frac{1}{4 \lambda \cdot \lambdabar}
	\left( \lambdabar^a \frac{\partial}{\partial \lambdabar_c}
	+ \lambdabar^c \frac{\partial}{\partial \lambdabar_a}
	- \lambda^a \frac{\partial}{\partial \lambda_c}
	- \lambda^c \frac{\partial}{\partial \lambda_a} \right)
	\left( \zeta^b \frac{\partial}{\partial \zeta^c}
	+ \zeta_c \frac{\partial}{\partial \zeta_b} \right) \bigg].
\end{align}
Its helicity components have been reported in eq.~\eqref{eq:K:2pt}.

For functions of the two momenta $p_1^\mu$ and $p_2^\mu$ defined in eq.~\eqref{eq:p1:p2:lambda}, the algebra is slightly more complicated but the logic is the same. One can show that derivatives with respect to $p_1$ and $p_2$ satisfy
\begin{align}
	\frac{\partial}{\partial p_{1\mu}}
	= \tfrac{1}{2} \sigma^\mu_{ab} \bigg[ &
	\frac{1}{\lambda \cdot \lambdabar}
	\left( \lambdabar^a \frac{\partial}{\partial \lambdabar_b}
	- \lambda^a \frac{\partial}{\partial \lambda_b} \right)
	-\frac{\lambda^a \lambda^b - \lambdabar^a \lambdabar^b}
	{(\lambda \cdot \lambdabar)^2} 
	\frac{1 - x + y}{1 - x - y}\, \H_0
	\nonumber \\
	& - 2 \frac{p_1^{ab}}{(\lambda \cdot \lambdabar)^2}
	\frac{1}{1 - x - y}
	\left( (1-x)^2 \frac{\partial}{\partial x}
	+ y^2 \frac{\partial}{\partial y} \right)
	\\
	& + 2 \frac{p_2^{ab}}{(\lambda \cdot \lambdabar)^2}
	\frac{1}{1 - x - y}
	\left( x (1 - x) \frac{\partial}{\partial x}
	+ y (1 - y) \frac{\partial}{\partial y} \right) \bigg],
	\nonumber \\
	\frac{\partial}{\partial p_{2\mu}}
	= \tfrac{1}{2}  \sigma^\mu_{ab} \bigg[ &
	\frac{1}{\lambda \cdot \lambdabar}
	\left( \lambdabar^a \frac{\partial}{\partial \lambdabar_b}
	- \lambda^a \frac{\partial}{\partial \lambda_b} \right)
	+ \frac{\lambda^a \lambda^b - \lambdabar^a \lambdabar^b}
	{(\lambda \cdot \lambdabar)^2}
	\frac{1 + x - y}{1 - x - y}\, \H_0
	\nonumber \\
	& + 2 \frac{p_1^{ab}}{(\lambda \cdot \lambdabar)^2}
	\frac{1}{1 - x - y}
	\left( x (1 - x) \frac{\partial}{\partial x}
	+ y (1 - y) \frac{\partial}{\partial y} \right)
	\\
	& - 2 \frac{p_2^{ab}}{(\lambda \cdot \lambdabar)^2}
	\frac{1}{1 - x - y}
	\left( x^2 \frac{\partial}{\partial x}
	+ (1 - y)^2 \frac{\partial}{\partial y} \right) \bigg],
	\nonumber 
\end{align}
where we have used the operator $\H_0$ defined in eq.~\eqref{eq:H}.
They can equivalently be rewritten in the helicity basis as
\begin{align}
	\frac{\partial}{\partial p_{1\mu}}
	= \frac{\sigma^\mu_{ab}}{2 (\lambda \cdot \lambdabar)^2}
	\bigg[ &
	-2 \lambda^a \lambdabar^b \left( \D
	+ (1 - x) \frac{\partial}{\partial x}
	- y \frac{\partial}{\partial y} \right)
	\nonumber \\
	& + \lambdabar^a \lambdabar^b \left( \H_+
	- (1 - x) \frac{\partial}{\partial x}
	- y \frac{\partial}{\partial y}
	+ \frac{1 - x + y}{1 - x - y} \H_0 \right)
	\label{eq:dp1:projection}
	\\
	& + \lambda^a \lambda^b \left( \H_- 
	- (1 - x) \frac{\partial}{\partial x}
	- y \frac{\partial}{\partial y}
	- \frac{1 - x + y}{1 - x - y} \H_0 \right) \bigg],
	\nonumber \\
	\frac{\partial}{\partial p_{2\mu}}
	= \frac{\sigma^\mu_{ab}}{2 (\lambda \cdot \lambdabar)^2}
	\bigg[ &
	-2 \lambda^a \lambdabar^b \left( \D
	- x \frac{\partial}{\partial x}
	+ (1 - y) \frac{\partial}{\partial y} \right)
	\nonumber \\
	& + \lambdabar^a \lambdabar^b \left( \H_+
	+ x \frac{\partial}{\partial x}
	+ (1 - y) \frac{\partial}{\partial y}
	- \frac{1 + x - y}{1 - x - y} \H_0 \right) 
	\label{eq:dp2:projection}
	\\
	& + \lambda^a \lambda^b \left( \H_- 
	+ x \frac{\partial}{\partial x}
	+ (1 - y) \frac{\partial}{\partial y}
	+ \frac{1 + x - y}{1 - x - y} \H_0 \right) \bigg].
	\nonumber
\end{align}
Note that when they are applied to functions that are independent of $x$ and $y$ and that carry zero total helicity, both differential operators reduce to $\partial/\partial p_\mu$ given in eq.~\eqref{eq:dp}. Moreover, one can form some linear combinations of the two operators that do not involve derivatives with respect to $x$ or $y$, and recover in this way the generators of scale and Lorentz transformations:
\begin{equation}
	\sum_{i = 1}^2 \, p_i^\mu \frac{\partial}{\partial p_i^\mu} = \D,
	\qquad\qquad
	\sum_{i = 1}^2 \left[
	p_{i\nu} \frac{\partial}{\partial p_i^\mu}
	- p_{i\mu} \frac{\partial}{\partial p_i^\nu} \right]
	= \M_{\mu\nu}.
	\label{eq:D:M:p1p2}
\end{equation}
This shows that the 3-point correlation function obeys the simple Ward identities \eqref{eq:WI:Lorentz:3pt} and \eqref{eq:WI:scale:3pt} that are independent of $x$ or $y$, as expected since the latter are dimensionless scalars.

On the contrary, the Ward identity for special conformal transformations involves derivatives with respect to $x$ and $y$.
If we denote with $\k^\mu$ the part of the differential operator~\eqref{eq:K:3pt:p} that is second-order in derivatives of $p_i$, we can write
\begin{equation}
	\k^\mu \equiv \sum_{i = 1}^2 \left[ \frac{1}{2} p_i^\mu 
	\frac{\partial^2}{\partial p_i^\nu \partial p_{i\nu}}
	- p_i^\nu \frac{\partial^2}{\partial p_i^\nu \partial p_{i\mu}} \right]
	= \frac{\sigma^\mu_{ab}}{2 (\lambda \cdot \lambdabar)^2}
	\left( \lambda^a \lambdabar^b \k_0
	+ \lambdabar^a \lambdabar^b \k_+
	+ \lambda^a \lambda^b \k_- \right),
\end{equation}
and we find
\begin{align}
	\k_0 &= 2 \A^\one_{3, 3; 0, 0}
	+ \frac{1}{2} ( \H_+ \H_- + \H_- \H_+ )
	+ \D (\D - 3)
	- \frac{(1 - x + y) (1 + x - y)}{(1 - x - y)^2} \H_0^2,
	\nonumber \\
	\k_+ &= \A^\two_{3,3}
	- \H_+ (\D + \H_0 - 1) 
	\nonumber \\
	& \quad
	- \frac{1}{1 - x - y} \left[
	2 x (1 - x) \frac{\partial}{\partial x}
	- 2 y (1 - y) \frac{\partial}{\partial y}
	+ (x - y) (2 - \H_0) \right] \H_0
	\nonumber \\
	\k_- &= \A^\two_{3,3}
	- \H_- (\D - \H_0 - 1) 
	\nonumber \\
	& \quad
	+ \frac{1}{1 - x - y} \left[
	2 x (1 - x) \frac{\partial}{\partial x}
	- 2 y (1 - y) \frac{\partial}{\partial y}
	+ (x - y) (2 + \H_0) \right] \H_0.
\end{align}
$\A^\one$ and $\A^\two$ are differential operators in $x$ and $y$, corresponding to the definitions~\eqref{eq:A:1} and \eqref{eq:A:2} but with special values of the parameters $\Delta_1 = \Delta_2 = 3$ and $\Delta = m = 0$. When combined with the derivatives in eqs.~\eqref{eq:dp1:projection} and \eqref{eq:dp2:projection}, we recover the operators \eqref{eq:K:3pt}.

The computations reported in this appendix can all be performed by hand, but the proliferation of spinors and indices can easily lead to errors.
For this reason, a computer algebra implementation of this formalism has been developed and used. It is made freely available in the form of a \texttt{Mathematica} notebook at the following address:

\begin{center}
	\url{https://github.com/gillioz/Conformal_spinors}
\end{center}


\section{Generalized hypergeometric series}
\label{sec:AppellF4}

This appendix discusses in some detail the algebra of partial differential equations generated by the operators $\A^\one$ and $\A^\two$ defined in eqs.~\eqref{eq:A:1} and \eqref{eq:A:2}, and whose indices we omit for compactness of the notation.
For simplicity, we will also use
\begin{align}
	\alpha &= \frac{6 - \Delta_1 - \Delta_2 - \Delta + j}{2}, &
	\gamma_1 &= \frac{5}{2} - \Delta_1,
	\nonumber \\
	\beta &= \frac{3 - \Delta_1 - \Delta_2 + \Delta + j}{2}, &
	\gamma_2 &= \frac{5}{2} - \Delta_2.
\end{align}
The first observation is that, when acting on a function of the two variables
\begin{equation}
	w_1 = x (1-y),
	\qquad\qquad
	w_2 = (1-x) y,
	\label{eq:w1w2}
\end{equation}
the differential operators $\A^\one$ and $\A^\two$ satisfy
\begin{align}
	\frac{1}{2} \bigg( \A^\one &
	+ \frac{1 - x + y}{1 - x -y} \A^\two \bigg)
	f(w_1, w_2)
	\nonumber \\
	= \bigg[ &
	w_1 (1 - w_1) \frac{\partial^2}{\partial w_1^2}
	- 2 w_1 w_2 \frac{\partial^2}{\partial w_1 \partial w_2}
	- w_2^2 \frac{\partial^2}{\partial w_2^2}
	\\
	& + \left( \gamma_1 - (\alpha + \beta + 1) w_1 \right)
	\frac{\partial}{\partial w_1}
	- (\alpha + \beta + 1) w_2 \frac{\partial}{\partial w_2}
	- \alpha \beta \bigg] f(w_1, w_2),
	\nonumber \\
	\frac{1}{2} \bigg( \A^\one &
	- \frac{1 + x - y}{1 - x -y} \A^\two \bigg)
	f(w_1, w_2)
	\nonumber \\
	= \bigg[ &
	w_2 (1 - w_2) \frac{\partial^2}{\partial w_2^2}
	- 2 w_1 w_2 \frac{\partial^2}{\partial w_1 \partial w_2}
	- w_1^2 \frac{\partial^2}{\partial w_1^2}
	\\
	& + \left( \gamma_2 - (\alpha + \beta + 1) w_2 \right)
	\frac{\partial}{\partial w_2}
	- (\alpha + \beta + 1) w_1 \frac{\partial}{\partial w_1}
	- \alpha \beta \bigg] f(w_1, w_2).
	\nonumber
\end{align}
This is the system of Appell $F_4$ differential equations in its standard form. It is solved by the double hypergeometric series
\begin{equation}
	f(w_1, w_2)
	= \sum_{i,j = 0}^\infty
	\frac{(\alpha)_{i+j} (\beta)_{i+j}}
	{i! j! (\gamma_1)_i (\gamma_2)_j}
	w_1^i w_2^j
	\equiv F_4(\alpha, \beta, \gamma_1, \gamma_2; w_1, w_2).
	\label{eq:F4:def}
\end{equation}
Alternatively, if one did not know about the Appell $F_4$ series, a solution to this system of differential equations could be obtained in the following way. Note that $\A^\one$ and $\A^\two$ look very much like the sum and difference of the differential operators generating the ordinary hypergeometric series in $x$ and $y$ respectively: if one denotes with
\begin{equation}
	\Q(\alpha, \beta, \gamma; x)
	= x (1-x) \frac{\partial^2}{\partial x^2}
	+ \left[ \gamma - (\alpha + \beta + 1) x \right]
	\frac{\partial}{\partial x}
	- \alpha \beta
\end{equation}
the operator that annihilate the ordinary hypergeometric function,
\begin{equation}
	\Q(\alpha, \beta, \gamma; x) \,
	{}_2F_1(\alpha, \beta, \gamma; x) = 0,
\end{equation}
then we have
\begin{align}
	\A^\one &= \Q(\alpha, \beta, \gamma_1; x)
	+ \Q(\alpha, \beta, \gamma_2; y)
	\nonumber \\
	& \quad
	- (\alpha + \beta - \gamma_1 - \gamma_2 + 1)
	\left[ \frac{1 - x + y}{1 - x - y} \frac{\partial}{\partial x}
	+ \frac{1 + x - y}{1 - x - y} \frac{\partial}{\partial y} \right]
\end{align}
and
\begin{equation}
	\A^\two = \Q(\alpha, \beta, \gamma_1; x)
	- \Q(\alpha, \beta, \gamma_2; y)
	+ (\alpha + \beta - \gamma_1 - \gamma_2 + 1)
	\left[ x \frac{\partial}{\partial x}
	- y \frac{\partial}{\partial y} \right].
\end{equation}
If it were not for the last term that involves both $x$ and $y$, these operators would simply generate decoupled systems of hypergeometric equations in $x$ and $y$.
Therefore, this suggests looking for a solution that can be written as a product of ${}_2F_1$ functions in $x$ and $y$.
If we introduce
\begin{equation}
	f_r(x, y) = x^r y^r
	{}_2F_1(\alpha + r, \beta + r, \gamma_1 + r; x)
	{}_2F_1(\alpha + r, \beta + r, \gamma_2 + r; y),
\end{equation}
then we can use properties of the ${}_2F_1$ function to show that
\begin{align}
	\A^\one f_r(x, y)
	&= (\alpha + r) (\alpha + \beta - \gamma_1 - \gamma_2 + 1 + r)
	f^\one_r(x,y)
	\nonumber \\
	& \quad
	- \frac{r (\gamma_1 + r - 1) (\gamma_2 + r - 1)}{\beta + r - 1}
	f^\one_{r-1}(x,y), 
	\label{eq:A:f:1}
	\\
	\A^\two f_r(x, y)
	&= (\alpha + r) (\alpha + \beta - \gamma_1 - \gamma_2 + 1 + r)
	f^\two_r(x,y)
	\nonumber \\
	& \quad
	- \frac{r (\gamma_1 + r - 1) (\gamma_2 + r - 1)}{\beta + r - 1}
	f^\two_{r-1}(x,y), 
	\label{eq:A:f:2}
\end{align}
where we have denoted
\begin{align}
	f^\one_r(x,y) = x^r y^r \bigg[ &
	\frac{2}{1 - x - y}
	{}_2F_1(\alpha + r, \beta + r, \gamma_1 + r; x)
	{}_2F_1(\alpha + r, \beta + r, \gamma_2 + r; y)
	\nonumber \\
	& - \frac{1 - x + y}{1 - x - y}
	{}_2F_1(\alpha + r + 1, \beta + r, \gamma_1 + r; x)
	{}_2F_1(\alpha + r, \beta + r, \gamma_2 + r; y)
	\nonumber \\
	& - \frac{1 + x - y}{1 - x - y}
	{}_2F_1(\alpha + r, \beta + r, \gamma_1 + r; x)
	{}_2F_1(\alpha + r + 1, \beta + r, \gamma_2 + r; y) \bigg]
	\nonumber \\
	f^\two_r(x,y) = x^r y^r \bigg[ &
	{}_2F_1(\alpha + r + 1, \beta + r, \gamma_1 + r; x)
	{}_2F_1(\alpha + r, \beta + r, \gamma_2 + r; y)
	\nonumber \\
	& - {}_2F_1(\alpha + r, \beta + r, \gamma_1 + r; x)
	{}_2F_1(\alpha + r + 1, \beta + r, \gamma_2 + r; y) \bigg].
\end{align}
Since eqs.~\eqref{eq:A:f:1} and \eqref{eq:A:f:2} are differences of consecutive terms in $r$, the infinite sum
\begin{equation}
	f(x,y) = \sum_{r = 0}^\infty 
	\frac{(\alpha)_r (\beta)_r
	(\alpha + \beta - \gamma_1 - \gamma_2 + 1)_r}
	{r! (\gamma_1)_r (\gamma_2)_r}
	f_r(x,y),
	\label{eq:F4:series:alt}
\end{equation}
satisfies
\begin{equation}
	\A^\one f(x,y) = \A^\two f(x,y) = 0.
\end{equation}
The series~\eqref{eq:F4:series:alt} is in fact a well-known representation of the Appell $F_4$ function, completely equivalent to the definition \eqref{eq:F4:def}.

The reason why we discuss this representation here is that it is a convenient starting point to solve the recursion relation \eqref{eq:Fjm:ladder}.
Upon acting on the series~\eqref{eq:F4:series:alt} with $\A^\two$, one encounters a generalization of that series, which we denote
\begin{align}
	f^{(a,b)}(x,y)
	= \sum_{r = 0}^\infty &
	\frac{(\alpha)_r (\beta)_r
	(\alpha + \beta - \gamma_1 - \gamma_2 + 1)_r}
	{r! (\gamma_1 + a)_r (\gamma_2 + b)_r}
	x^{a + r} y^{b + r}
	\nonumber \\
	& \times
	{}_2F_1(\alpha + r, \beta + r, \gamma_1 + a + r; x)
	{}_2F_1(\alpha + r, \beta + r, \gamma_2 + b + r; y).
\end{align}
This is an infinite series in $x$ and $y$ that begins at order $x^a y^b$. The case $a = b = 0$ corresponds to the Appell $F_4$ series, $f^{(0,0)}(x,y) = f(x, y)$.
The general solution to the recursion equation could be guessed after examining the first few steps, and then verified to hold at higher orders.
It is given by the linear combination
\begin{equation}
	f(x,y) = \sum_{n = 0}^{j - m} 
	(1 - x - y)^{j - n}
	\sum_{i = 0}^n
	\chi^{(n,i)}
	f^{(n-i, i)}(x,y),
	\label{eq:fjm:linearcombination}
\end{equation}
with coefficients
\begin{align}
	\chi^{(n,i)} &= \frac{(-1)^i}{i! (n - i)!}
	\frac{4^{j - m} (m + 1)_{j - m}}
	{(2m + 1)_{j - m - n}}
	\frac{(\gamma_1 - \beta)_{n-i} (\gamma_2 - \beta)_i}
	{(\gamma_1)_{n-i} (\gamma_2)_i
	(\gamma_1 + \gamma_2 - 2\beta)_{j - m}}
	\nonumber \\
	& \quad \times
	\sum_{k = 0}^{j - m - n}
	\frac{(-1)^k
	\left( m + \frac{1}{2} \right)_{j - m - n - k}
	\left( m + \frac{1}{2} \right)_k}
	{k! (j - m - n - k)!}
	(\gamma_1 - \alpha)_{j - m - i - k}
	(\gamma_2 - \alpha)_{i + k}.
\end{align}
These are precisely the results reported in eqs.~\eqref{eq:fjm}, \eqref{eq:fab}, and \eqref{eq:chi}.

This construction so far is specific to the solution of the system of differential equations that is analytic in $x$ and $y$ around $(x, y) = (0, 0)$.
However, the other solutions can easily be extracted from it, noting that the differential operators $\A^\one$ and $\A^\two$ obey
\begin{equation}
\begin{aligned}
	\A^\one_{\Delta_1, \Delta_2, \Delta, m}
	\left[ w_1^{\Delta_1 - 3/2} f(x,y) \right]
	&= w_1^{\Delta_1 - 3/2}
	\A^\one_{3 - \Delta_1, \Delta_2, \Delta, m} f(x,y)
	\\
	\A^\two_{\Delta_1, \Delta_2}
	\left[ w_1^{\Delta_1 - 3/2} f(x,y) \right]
	&= w_1^{\Delta_1 - 3/2}
	\A^\two_{3 - \Delta_1, \Delta_2} f(x,y)
\end{aligned}
\end{equation}
and
\begin{equation}
\begin{aligned}
	\A^\one_{\Delta_1, \Delta_2, \Delta, m}
	\left[ w_2^{\Delta_2 - 3/2} f(x,y) \right]
	&= w_2^{\Delta_2 - 3/2}
	\A^\one_{\Delta_1, 3 - \Delta_2, \Delta, m} f(x,y)
	\\
	\A^\two_{\Delta_1, \Delta_2}
	\left[ w_2^{\Delta_2 - 3/2} f(x,y) \right]
	&= w_2^{\Delta_2 - 3/2}
	\A^\two_{\Delta_1, 3 - \Delta_2} f(x,y)
\end{aligned}
\end{equation}
This explains immediately why the four functions \eqref{eq:4solutions} are related by the ``shadow transform'' $\Delta_i \leftrightarrow 3 - \Delta_i$ for $i = 1, 2$. But since the recursion relation \eqref{eq:Fjm:ladder} also simply depends on $\A^\two$, the construction of the solution follows the same rule, and the result is directly given by the linear combination \eqref{eq:fjm:linearcombination} in which the transformation $\Delta_i \leftrightarrow 3 - \Delta_i$ is applied both to the functions $f^{(n-i,i)}(x,y)$ and to the coefficients $\chi^{(n,i)}$.

\paragraph{Colinear limit.}

Note that in the colinear limit in which $1 - x - y = 0$, the function $f(x,y)$ vanishes unless $m = 0$. Moreover, at $m = 0$, only the terms with $n = j$ contribute to the sum \eqref{eq:fjm:linearcombination}. This means that we have
\begin{equation}
	f(x,y)
	= \sum_{i = 0}^j
	\chi^{(j,i)}
	f^{(j-i, i)}(x,y),
	\label{eq:fjm:colinear}
\end{equation}
with coefficients taking the simpler form
\begin{equation}
	\chi^{(j,i)} = \frac{(-1)^i 4^j j!}{i! (j - i)!}
	\frac{(\gamma_1 - \alpha)_{j-i} (\gamma_1 - \beta)_{j-i}
	(\gamma_2 - \alpha)_{i} (\gamma_2 - \beta)_i}
	{(\gamma_1)_{j-i} (\gamma_2)_i
	(\gamma_1 + \gamma_2 - 2\beta)_j}.
	\label{eq:chi:colinear}
\end{equation}

\paragraph{Light-cone limit.}

Finally, it should be noted that our formulation of the solution is particularly well-suited to examine either of the limits $x \to 0$ or $y \to 0$, corresponding respectively to $p_1^2 \to 0$ or $p_2^2 \to 0$. Let us describe the latter limit, $y \to 0$.
Since the function $f^{(a,b)}(x,y)$ is a series of non-negative powers of $y$ starting with $y^b$, only the functions with $b = 0$ contribute to that limit. Moreover, only the leading term with $r = 0$ contributes in the sum, so that we find
\begin{equation}
	F_{j,m}(x,0) = \sum_{n = 0}^{j - m}
	(1 - x)^{j - n}
	\chi^{(n,0)}
	x^n {}_2F_1(\alpha, \beta, \gamma_1 + n; x).
\end{equation}
Using the properties of the coefficient $\chi^{(n,0)}$, this can be rewritten
\begin{align}
	F_{j,m}(x,0) = \sum_{n = 0}^{j - m} &
	\frac{(-1)^n 4^{j-m} j!
	\left( m + \frac{1}{2} \right)_{j - m - n}
	(\Delta - j - 1 + n)_{j - m -n }
	\left( \frac{\Delta - j - 1 - \Delta_1 + \Delta_2}{2} \right)_n}
	{m! n! (j - m - n)! 
	(2m + 1)_{j - m - n}
	(\Delta + m - 1)_{j-m}}
	\nonumber \\
	& \times
	(1 - x)^{j - n}
	{}_2F_1\left( \tfrac{6 - \Delta_1 - \Delta_2 - \Delta + j}{2} - n,
	\tfrac{3 - \Delta_1 - \Delta_2 + \Delta + j}{2},
	\tfrac{5}{2} - \Delta_1; x \right).
\end{align}
This matches the result obtained in arbitrary space-time dimension in ref.~\cite{Gillioz:2020wgw}.

\bibliography{Bibliography}

\providecommand{\href}[2]{#2}\begingroup\raggedright\begin{thebibliography}{10}

\bibitem{Osborn:1993cr}
H.~Osborn and A.~C. Petkou, ``{Implications of conformal invariance in field
  theories for general dimensions},''
  \href{http://dx.doi.org/10.1006/aphy.1994.1045}{{\em Annals Phys.} {\bfseries
  231} (1994) 311--362}, \href{http://arxiv.org/abs/hep-th/9307010}{{\ttfamily
  arXiv:hep-th/9307010}}.

\bibitem{Costa:2011mg}
M.~S. Costa, J.~Penedones, D.~Poland, and S.~Rychkov, ``{Spinning Conformal
  Correlators},'' \href{http://dx.doi.org/10.1007/JHEP11(2011)071}{{\em JHEP}
  {\bfseries 11} (2011) 071}, \href{http://arxiv.org/abs/1107.3554}{{\ttfamily
  arXiv:1107.3554 [hep-th]}}.

\bibitem{Costa:2014rya}
M.~S. Costa and T.~Hansen, ``{Conformal correlators of mixed-symmetry
  tensors},'' \href{http://dx.doi.org/10.1007/JHEP02(2015)151}{{\em JHEP}
  {\bfseries 02} (2015) 151}, \href{http://arxiv.org/abs/1411.7351}{{\ttfamily
  arXiv:1411.7351 [hep-th]}}.

\bibitem{Coriano:2013jba}
C.~Coriano, L.~Delle~Rose, E.~Mottola, and M.~Serino, ``{Solving the Conformal
  Constraints for Scalar Operators in Momentum Space and the Evaluation of
  Feynman's Master Integrals},''
  \href{http://dx.doi.org/10.1007/JHEP07(2013)011}{{\em JHEP} {\bfseries 07}
  (2013) 011}, \href{http://arxiv.org/abs/1304.6944}{{\ttfamily arXiv:1304.6944
  [hep-th]}}.

\bibitem{Bzowski:2013sza}
A.~Bzowski, P.~McFadden, and K.~Skenderis, ``{Implications of conformal
  invariance in momentum space},''
  \href{http://dx.doi.org/10.1007/JHEP03(2014)111}{{\em JHEP} {\bfseries 03}
  (2014) 111}, \href{http://arxiv.org/abs/1304.7760}{{\ttfamily arXiv:1304.7760
  [hep-th]}}.

\bibitem{Caron-Huot:2021kjy}
S.~Caron-Huot and Y.-Z. Li, ``{Helicity basis for three-dimensional conformal
  field theory},'' \href{http://dx.doi.org/10.1007/JHEP06(2021)041}{{\em JHEP}
  {\bfseries 06} (2021) 041}, \href{http://arxiv.org/abs/2102.08160}{{\ttfamily
  arXiv:2102.08160 [hep-th]}}.

\bibitem{Jain:2021vrv}
S.~Jain, R.~R. John, A.~Mehta, A.~A. Nizami, and A.~Suresh, ``{Higher spin
  3-point functions in 3d CFT using spinor-helicity variables},''
  \href{http://dx.doi.org/10.1007/JHEP09(2021)041}{{\em JHEP} {\bfseries 09}
  (2021) 041}, \href{http://arxiv.org/abs/2106.00016}{{\ttfamily
  arXiv:2106.00016 [hep-th]}}.

\bibitem{Gillioz:2018mto}
M.~Gillioz, ``{Momentum-space conformal blocks on the light cone},''
  \href{http://dx.doi.org/10.1007/JHEP10(2018)125}{{\em JHEP} {\bfseries 10}
  (2018) 125}, \href{http://arxiv.org/abs/1807.07003}{{\ttfamily
  arXiv:1807.07003 [hep-th]}}.

\bibitem{Gillioz:2019lgs}
M.~Gillioz, ``{Conformal 3-point functions and the Lorentzian OPE in momentum
  space},'' \href{http://dx.doi.org/10.1007/s00220-020-03836-8}{{\em Commun.
  Math. Phys.} {\bfseries 379} no.~1, (2020) 227--259},
  \href{http://arxiv.org/abs/1909.00878}{{\ttfamily arXiv:1909.00878
  [hep-th]}}.

\bibitem{Mack:1969dg}
G.~Mack and I.~Todorov, ``{Irreducibility of the ladder representations of
  u(2,2) when restricted to the Poincaré subgroup},''
  \href{http://dx.doi.org/10.1063/1.1664804}{{\em J. Math. Phys.} {\bfseries
  10} (1969) 2078--2085}.

\bibitem{Mack:1975je}
G.~Mack, ``{All unitary ray representations of the conformal group SU(2,2) with
  positive energy},'' \href{http://dx.doi.org/10.1007/BF01613145}{{\em Commun.
  Math. Phys.} {\bfseries 55} (1977) 1}.

\bibitem{Ferrara:1974pt}
S.~Ferrara, R.~Gatto, and A.~F. Grillo, ``{Positivity Restrictions on Anomalous
  Dimensions},'' \href{http://dx.doi.org/10.1103/PhysRevD.9.3564}{{\em Phys.
  Rev. D} {\bfseries 9} (1974) 3564}.

\bibitem{Minwalla:1997ka}
S.~Minwalla, ``{Restrictions imposed by superconformal invariance on quantum
  field theories},'' \href{http://dx.doi.org/10.4310/ATMP.1998.v2.n4.a4}{{\em
  Adv. Theor. Math. Phys.} {\bfseries 2} (1998) 783--851},
  \href{http://arxiv.org/abs/hep-th/9712074}{{\ttfamily arXiv:hep-th/9712074}}.

\bibitem{Grinstein:2008qk}
B.~Grinstein, K.~A. Intriligator, and I.~Z. Rothstein, ``{Comments on
  Unparticles},'' \href{http://dx.doi.org/10.1016/j.physletb.2008.03.020}{{\em
  Phys. Lett. B} {\bfseries 662} (2008) 367--374},
  \href{http://arxiv.org/abs/0801.1140}{{\ttfamily arXiv:0801.1140 [hep-ph]}}.

\bibitem{Arkani-Hamed:2017jhn}
N.~Arkani-Hamed, T.-C. Huang, and Y.-t. Huang, ``{Scattering amplitudes for all
  masses and spins},'' \href{http://dx.doi.org/10.1007/JHEP11(2021)070}{{\em
  JHEP} {\bfseries 11} (2021) 070},
  \href{http://arxiv.org/abs/1709.04891}{{\ttfamily arXiv:1709.04891
  [hep-th]}}.

\bibitem{Dobrev:1975ru}
V.~K. Dobrev, V.~B. Petkova, S.~G. Petrova, and I.~T. Todorov, ``{Dynamical
  Derivation of Vacuum Operator Product Expansion in Euclidean Conformal
  Quantum Field Theory},''
  \href{http://dx.doi.org/10.1103/PhysRevD.13.887}{{\em Phys. Rev. D}
  {\bfseries 13} (1976) 887}.

\bibitem{Maldacena:2011nz}
J.~M. Maldacena and G.~L. Pimentel, ``{On graviton non-Gaussianities during
  inflation},'' \href{http://dx.doi.org/10.1007/JHEP09(2011)045}{{\em JHEP}
  {\bfseries 09} (2011) 045}, \href{http://arxiv.org/abs/1104.2846}{{\ttfamily
  arXiv:1104.2846 [hep-th]}}.

\bibitem{Erramilli:2019njx}
R.~S. Erramilli, L.~V. Iliesiu, and P.~Kravchuk, ``{Recursion relation for
  general 3d blocks},'' \href{http://dx.doi.org/10.1007/JHEP12(2019)116}{{\em
  JHEP} {\bfseries 12} (2019) 116},
  \href{http://arxiv.org/abs/1907.11247}{{\ttfamily arXiv:1907.11247
  [hep-th]}}.

\bibitem{Hogervorst:2016itc}
M.~Hogervorst, M.~Paulos, and A.~Vichi, ``{The ABC (in any D) of Logarithmic
  CFT},'' \href{http://dx.doi.org/10.1007/JHEP10(2017)201}{{\em JHEP}
  {\bfseries 10} (2017) 201}, \href{http://arxiv.org/abs/1605.03959}{{\ttfamily
  arXiv:1605.03959 [hep-th]}}.

\bibitem{Henning:2019mcv}
B.~Henning and T.~Melia, ``{Conformal-helicity duality \& the Hilbert space of
  free CFTs},'' \href{http://arxiv.org/abs/1902.06747}{{\ttfamily
  arXiv:1902.06747 [hep-th]}}.

\bibitem{Henning:2019enq}
B.~Henning and T.~Melia, ``{Constructing effective field theories via their
  harmonics},'' \href{http://dx.doi.org/10.1103/PhysRevD.100.016015}{{\em Phys.
  Rev. D} {\bfseries 100} no.~1, (2019) 016015},
  \href{http://arxiv.org/abs/1902.06754}{{\ttfamily arXiv:1902.06754
  [hep-ph]}}.

\bibitem{Exton:1995}
H.~Exton, ``{On the system of partial differential equations associated with
  Appell's function F4},''
  \href{http://dx.doi.org/10.1088/0305-4470/28/3/017}{{\em Phys. A: Math. Gen.}
  {\bfseries 28} (1995) 631}.

\bibitem{Gillioz:2020wgw}
M.~Gillioz, ``{Conformal partial waves in momentum space},''
  \href{http://dx.doi.org/10.21468/SciPostPhys.10.4.081}{{\em SciPost Phys.}
  {\bfseries 10} no.~4, (2021) 081},
  \href{http://arxiv.org/abs/2012.09825}{{\ttfamily arXiv:2012.09825
  [hep-th]}}.

\bibitem{Gillioz:2021sce}
M.~Gillioz, ``{From Schwinger to Wightman: all conformal 3-point functions in
  momentum space},'' \href{http://arxiv.org/abs/2109.15140}{{\ttfamily
  arXiv:2109.15140 [hep-th]}}.

\bibitem{Gillioz:2019iye}
M.~Gillioz, X.~Lu, M.~A. Luty, and G.~Mikaberidze, ``{Convergent Momentum-Space
  OPE and Bootstrap Equations in Conformal Field Theory},''
  \href{http://dx.doi.org/10.1007/JHEP03(2020)102}{{\em JHEP} {\bfseries 03}
  (2020) 102}, \href{http://arxiv.org/abs/1912.05550}{{\ttfamily
  arXiv:1912.05550 [hep-th]}}.

\bibitem{Bautista:2019qxj}
T.~Bautista and H.~Godazgar, ``{Lorentzian CFT 3-point functions in momentum
  space},'' \href{http://dx.doi.org/10.1007/JHEP01(2020)142}{{\em JHEP}
  {\bfseries 01} (2020) 142}, \href{http://arxiv.org/abs/1908.04733}{{\ttfamily
  arXiv:1908.04733 [hep-th]}}.

\bibitem{Costa:2014kfa}
M.~S. Costa, V.~Gon\c{c}alves, and J.~Penedones, ``{Spinning AdS
  Propagators},'' \href{http://dx.doi.org/10.1007/JHEP09(2014)064}{{\em JHEP}
  {\bfseries 09} (2014) 064}, \href{http://arxiv.org/abs/1404.5625}{{\ttfamily
  arXiv:1404.5625 [hep-th]}}.

\bibitem{Costa:2018mcg}
M.~S. Costa and T.~Hansen, ``{AdS Weight Shifting Operators},''
  \href{http://dx.doi.org/10.1007/JHEP09(2018)040}{{\em JHEP} {\bfseries 09}
  (2018) 040}, \href{http://arxiv.org/abs/1805.01492}{{\ttfamily
  arXiv:1805.01492 [hep-th]}}.

\bibitem{Meltzer:2019nbs}
D.~Meltzer, E.~Perlmutter, and A.~Sivaramakrishnan, ``{Unitarity Methods in
  AdS/CFT},'' \href{http://dx.doi.org/10.1007/JHEP03(2020)061}{{\em JHEP}
  {\bfseries 03} (2020) 061}, \href{http://arxiv.org/abs/1912.09521}{{\ttfamily
  arXiv:1912.09521 [hep-th]}}.

\bibitem{Fortin:2019dnq}
J.-F. Fortin and W.~Skiba, ``{New methods for conformal correlation
  functions},'' \href{http://dx.doi.org/10.1007/JHEP06(2020)028}{{\em JHEP}
  {\bfseries 06} (2020) 028}, \href{http://arxiv.org/abs/1905.00434}{{\ttfamily
  arXiv:1905.00434 [hep-th]}}.

\bibitem{Sleight:2017fpc}
C.~Sleight and M.~Taronna, ``{Spinning Witten Diagrams},''
  \href{http://dx.doi.org/10.1007/JHEP06(2017)100}{{\em JHEP} {\bfseries 06}
  (2017) 100}, \href{http://arxiv.org/abs/1702.08619}{{\ttfamily
  arXiv:1702.08619 [hep-th]}}.

\bibitem{Isono:2017grm}
H.~Isono, ``{On conformal correlators and blocks with spinors in general
  dimensions},'' \href{http://dx.doi.org/10.1103/PhysRevD.96.065011}{{\em Phys.
  Rev. D} {\bfseries 96} no.~6, (2017) 065011},
  \href{http://arxiv.org/abs/1706.02835}{{\ttfamily arXiv:1706.02835
  [hep-th]}}.

\bibitem{Karateev:2017jgd}
D.~Karateev, P.~Kravchuk, and D.~Simmons-Duffin, ``{Weight Shifting Operators
  and Conformal Blocks},''
  \href{http://dx.doi.org/10.1007/JHEP02(2018)081}{{\em JHEP} {\bfseries 02}
  (2018) 081}, \href{http://arxiv.org/abs/1706.07813}{{\ttfamily
  arXiv:1706.07813 [hep-th]}}.

\bibitem{Karateev:2018oml}
D.~Karateev, P.~Kravchuk, and D.~Simmons-Duffin, ``{Harmonic Analysis and Mean
  Field Theory},'' \href{http://dx.doi.org/10.1007/JHEP10(2019)217}{{\em JHEP}
  {\bfseries 10} (2019) 217}, \href{http://arxiv.org/abs/1809.05111}{{\ttfamily
  arXiv:1809.05111 [hep-th]}}.

\bibitem{Kravchuk:2021kwe}
P.~Kravchuk, J.~Qiao, and S.~Rychkov, ``{Distributions in CFT. Part II.
  Minkowski space},'' \href{http://dx.doi.org/10.1007/JHEP08(2021)094}{{\em
  JHEP} {\bfseries 08} (2021) 094},
  \href{http://arxiv.org/abs/2104.02090}{{\ttfamily arXiv:2104.02090
  [hep-th]}}.

\bibitem{Rosenhaus:2018zqn}
V.~Rosenhaus, ``{Multipoint Conformal Blocks in the Comb Channel},''
  \href{http://dx.doi.org/10.1007/JHEP02(2019)142}{{\em JHEP} {\bfseries 02}
  (2019) 142}, \href{http://arxiv.org/abs/1810.03244}{{\ttfamily
  arXiv:1810.03244 [hep-th]}}.

\bibitem{Goncalves:2019znr}
V.~Gon\c{c}alves, R.~Pereira, and X.~Zhou, ``{$20'$ Five-Point Function from
  $AdS_5\times S^5$ Supergravity},''
  \href{http://dx.doi.org/10.1007/JHEP10(2019)247}{{\em JHEP} {\bfseries 10}
  (2019) 247}, \href{http://arxiv.org/abs/1906.05305}{{\ttfamily
  arXiv:1906.05305 [hep-th]}}.

\bibitem{Fortin:2019zkm}
J.-F. Fortin, W.~Ma, and W.~Skiba, ``{Higher-Point Conformal Blocks in the Comb
  Channel},'' \href{http://dx.doi.org/10.1007/JHEP07(2020)213}{{\em JHEP}
  {\bfseries 07} (2020) 213}, \href{http://arxiv.org/abs/1911.11046}{{\ttfamily
  arXiv:1911.11046 [hep-th]}}.

\bibitem{Bercini:2020msp}
C.~Bercini, V.~Gon\c{c}alves, and P.~Vieira, ``{Light-Cone Bootstrap of Higher
  Point Functions and Wilson Loop Duality},''
  \href{http://dx.doi.org/10.1103/PhysRevLett.126.121603}{{\em Phys. Rev.
  Lett.} {\bfseries 126} no.~12, (2021) 121603},
  \href{http://arxiv.org/abs/2008.10407}{{\ttfamily arXiv:2008.10407
  [hep-th]}}.

\bibitem{Poland:2021xjs}
D.~Poland and V.~Prilepina, ``{Recursion relations for 5-point conformal
  blocks},'' \href{http://dx.doi.org/10.1007/JHEP10(2021)160}{{\em JHEP}
  {\bfseries 10} (2021) 160}, \href{http://arxiv.org/abs/2103.12092}{{\ttfamily
  arXiv:2103.12092 [hep-th]}}.

\bibitem{Buric:2021ywo}
I.~Buric, S.~Lacroix, J.~A. Mann, L.~Quintavalle, and V.~Schomerus, ``{Gaudin
  models and multipoint conformal blocks: general theory},''
  \href{http://dx.doi.org/10.1007/JHEP10(2021)139}{{\em JHEP} {\bfseries 10}
  (2021) 139}, \href{http://arxiv.org/abs/2105.00021}{{\ttfamily
  arXiv:2105.00021 [hep-th]}}.

\bibitem{Buric:2021kgy}
I.~Buric, S.~Lacroix, J.~A. Mann, L.~Quintavalle, and V.~Schomerus, ``{Gaudin
  Models and Multipoint Conformal Blocks III: Comb channel coordinates and OPE
  factorisation},'' \href{http://arxiv.org/abs/2112.10827}{{\ttfamily
  arXiv:2112.10827 [hep-th]}}.

\bibitem{Antunes:2021kmm}
A.~Antunes, M.~S. Costa, V.~Goncalves, and J.~V. Boas, ``{Lightcone Bootstrap
  at higher points},'' \href{http://arxiv.org/abs/2111.05453}{{\ttfamily
  arXiv:2111.05453 [hep-th]}}.

\bibitem{Katz:2016hxp}
E.~Katz, Z.~U. Khandker, and M.~T. Walters, ``{A Conformal Truncation Framework
  for Infinite-Volume Dynamics},''
  \href{http://dx.doi.org/10.1007/JHEP07(2016)140}{{\em JHEP} {\bfseries 07}
  (2016) 140}, \href{http://arxiv.org/abs/1604.01766}{{\ttfamily
  arXiv:1604.01766 [hep-th]}}.

\bibitem{Anand:2020gnn}
N.~Anand, A.~L. Fitzpatrick, E.~Katz, Z.~U. Khandker, M.~T. Walters, and
  Y.~Xin, ``{Introduction to Lightcone Conformal Truncation: QFT Dynamics from
  CFT Data},'' \href{http://arxiv.org/abs/2005.13544}{{\ttfamily
  arXiv:2005.13544 [hep-th]}}.

\bibitem{Anand:2019lkt}
N.~Anand, Z.~U. Khandker, and M.~T. Walters, ``{Momentum space CFT correlators
  for Hamiltonian truncation},''
  \href{http://dx.doi.org/10.1007/JHEP10(2020)095}{{\em JHEP} {\bfseries 10}
  (2020) 095}, \href{http://arxiv.org/abs/1911.02573}{{\ttfamily
  arXiv:1911.02573 [hep-th]}}.

\bibitem{Arkani-Hamed:2018kmz}
N.~Arkani-Hamed, D.~Baumann, H.~Lee, and G.~L. Pimentel, ``{The Cosmological
  Bootstrap: Inflationary Correlators from Symmetries and Singularities},''
  \href{http://dx.doi.org/10.1007/JHEP04(2020)105}{{\em JHEP} {\bfseries 04}
  (2020) 105}, \href{http://arxiv.org/abs/1811.00024}{{\ttfamily
  arXiv:1811.00024 [hep-th]}}.

\bibitem{Baumann:2019oyu}
D.~Baumann, C.~Duaso~Pueyo, A.~Joyce, H.~Lee, and G.~L. Pimentel, ``{The
  cosmological bootstrap: weight-shifting operators and scalar seeds},''
  \href{http://dx.doi.org/10.1007/JHEP12(2020)204}{{\em JHEP} {\bfseries 12}
  (2020) 204}, \href{http://arxiv.org/abs/1910.14051}{{\ttfamily
  arXiv:1910.14051 [hep-th]}}.

\bibitem{Baumann:2020dch}
D.~Baumann, C.~Duaso~Pueyo, A.~Joyce, H.~Lee, and G.~L. Pimentel, ``{The
  Cosmological Bootstrap: Spinning Correlators from Symmetries and
  Factorization},'' \href{http://dx.doi.org/10.21468/SciPostPhys.11.3.071}{{\em
  SciPost Phys.} {\bfseries 11} (2021) 071},
  \href{http://arxiv.org/abs/2005.04234}{{\ttfamily arXiv:2005.04234
  [hep-th]}}.

\bibitem{Hogervorst:2021uvp}
M.~Hogervorst, J.~Penedones, and K.~S. Vaziri, ``{Towards the non-perturbative
  cosmological bootstrap},'' \href{http://arxiv.org/abs/2107.13871}{{\ttfamily
  arXiv:2107.13871 [hep-th]}}.

\bibitem{DiPietro:2021sjt}
L.~Di~Pietro, V.~Gorbenko, and S.~Komatsu, ``{Analyticity and unitarity for
  cosmological correlators},''
  \href{http://dx.doi.org/10.1007/JHEP03(2022)023}{{\em JHEP} {\bfseries 03}
  (2022) 023}, \href{http://arxiv.org/abs/2108.01695}{{\ttfamily
  arXiv:2108.01695 [hep-th]}}.

\bibitem{Pethybridge:2021rwf}
B.~Pethybridge and V.~Schaub, ``{Tensors and Spinors in de Sitter Space},''
  \href{http://arxiv.org/abs/2111.14899}{{\ttfamily arXiv:2111.14899
  [hep-th]}}.

\bibitem{Higuchi:2010xt}
A.~Higuchi, D.~Marolf, and I.~A. Morrison, ``{On the Equivalence between
  Euclidean and In-In Formalisms in de Sitter QFT},''
  \href{http://dx.doi.org/10.1103/PhysRevD.83.084029}{{\em Phys. Rev. D}
  {\bfseries 83} (2011) 084029},
  \href{http://arxiv.org/abs/1012.3415}{{\ttfamily arXiv:1012.3415 [gr-qc]}}.

\bibitem{Gorbenko:2019rza}
V.~Gorbenko and L.~Senatore, ``{$\lambda \phi^4$ in dS},''
  \href{http://arxiv.org/abs/1911.00022}{{\ttfamily arXiv:1911.00022
  [hep-th]}}.

\bibitem{Sleight:2020obc}
C.~Sleight and M.~Taronna, ``{From AdS to dS Exchanges: Spectral
  Representation, Mellin Amplitudes and Crossing},''
  \href{http://dx.doi.org/10.1103/PhysRevD.104.L081902}{{\em Phys. Rev. D}
  {\bfseries 104} (2021) 081902},
  \href{http://arxiv.org/abs/2007.09993}{{\ttfamily arXiv:2007.09993
  [hep-th]}}.

\bibitem{Sleight:2021plv}
C.~Sleight and M.~Taronna, ``{From dS to AdS and back},''
  \href{http://dx.doi.org/10.1007/JHEP12(2021)074}{{\em JHEP} {\bfseries 12}
  (2021) 074}, \href{http://arxiv.org/abs/2109.02725}{{\ttfamily
  arXiv:2109.02725 [hep-th]}}.

\end{thebibliography}\endgroup
\bibliographystyle{utphys}

\end{document}